\documentclass[twocolumn,aps,superscriptaddress]{revtex4-1}
\usepackage{palatino}
\usepackage[latin1]{inputenc}
\usepackage{epsf}
\usepackage{amsmath,amssymb}
\usepackage{latexsym}
\usepackage{calc}
\usepackage{color}
\usepackage{graphicx}

\newcommand{\ben}{\begin{equation}}
\newcommand{\een}{\end{equation}}
\newcommand{\bea}{\begin{eqnarray}}
\newcommand{\eea}{\end{eqnarray}}

\def\bra#1{\langle#1\vert}
\def\ket#1{\vert#1\rangle}

\def\sss{\scriptscriptstyle\rm}

\def\1s{_{1,\sss S}}
\def\2s{_{2,\sss S}}

\def\bq{{\bf q}}

\begin{document}
\title{Analysis of the Classical Trajectory Treatment of Photon Dynamics for Polaritonic Phenomena}
\author{Bart Rosenzweig}
\affiliation{Department of Mathematics and Statistics, Hunter College of the City University of New York, 695 Park Avenue, New York, New York 10065, USA}
\altaffiliation[Current Address: ]{Department of Mathematics, The Ohio State University, 231 W 18th Avenue, Columbus, OH 43210 }
\author{Norah M. Hoffmann}
\affiliation{Department of Chemistry, Columbia University, New York, New York 10027 USA}
\author{Lionel Lacombe}
\affiliation{Department of Physics, Rutgers University, Newark 07102, New Jersey USA}
\altaffiliation[Current address:]{Laboratoire des Solides Irradi{\'e}s, {\'E}cole Polytechnique, Institut Polytechnique de Paris, F-91128 Palaiseau, France}
\author{Neepa T. Maitra}
\affiliation{Department of Physics, Rutgers University, Newark 07102, New Jersey USA}
\email{neepa.maitra@rutgers.edu}
\date{\today}
\begin{abstract}
Simulating photon dynamics in strong light-matter coupling situations via classical trajectories  is proving to be powerful and practical. Here we analyze the performance of the approach through the lens of the exact factorization approach. Since the exact factorization enables a rigorous definition of the potentials driving the photonic motion it allows us to identify that the cause of the underestimation of photon number and intensities observed in earlier work is primarily due to an inadequate accounting of light-matter correlation in the classical Ehrenfest force rather than errors from treating the photons quasiclassically {\it per se}. The latter becomes problematic when the number of photons per mode begins to exceed a half. 
 \end{abstract}

\maketitle


\section{Introduction}
\label{sec:intro}

The impressive advances in experimental realizations of strong light-matter coupling induced by confinement
continue to open exciting new possibilities for the manipulation of matter~\cite{E16,HO20,GCE21,HWMB19}. This has led to the establishment of a new field of research in chemistry in the past decade, known as polaritonic chemistry, or molecular polaritonics.  The possible applications are ever-increasing and just within the last year we have seen experimental demonstrations of new phenomena in different directions, including  tunable self-assembled polaritons in microcavities~\cite{F21,Munkhbat2021}, solar light-harvesting enabled via cavity-coupling~\cite{Esteso2021}, and large enhancement of ferromagnetism from collective strong-coupling~\cite{Thomas2021}. In most cases, to achieve strong light-matter coupling a large number of molecules is required, exploiting the collective nature of the coupling that scales as the square-root of the number of molecules. The coupling strength also scales as the inverse square-root of the volume, so to reach strong-coupling with single molecules, a tighter confinement is required, and here plasmonic cavities have been a successful solution~\cite{Chikkaraddy2016,Santhosh2016,Ojambati2019,Park2019}. 

A key issue for plasmonic cavities is the large loss entailed from the electron oscillations in the metal, and experimental and theoretical works are on-going to deal with these, and even sometimes take advantage, e.g. ~\cite{SPGF20,TF21,Khurgin2015,UV20,DCY21,ASFVF20,FFSRVF20}.  Another challenge in the computational modeling of polaritonic systems in general is the efficient accounting of the many photon modes in the cavity that can influence the dynamics. 
In this work, we focus on this latter issue; in particular, on the treatment of photons via classical trajectory dynamics. 

When more than one cavity mode plays an important role, a full quantum treatment of the photonic degrees of freedom becomes challenging. Different approaches have been explored, including quantum electrodynamical density functional theory~\cite{T13,RFPATR14,RTFAR18}, cumulant expansions~\cite{SSF20}, tensor networks~\cite{PSCFG18}, and quasi-normal modes~\cite{FHDKBKR19} when accounting also for dissipation. Inspired by mixed quantum-classical methods used for electron-nuclear coupled motion, the multi-trajectory Ehrenfest (MTE) approach was proposed~\cite{HSRKA19, HSSRAK19}  and demonstrated to be an efficient and powerful approach. In MTE, the photonic system is represented by the displacement-field coordinate $q$ and its conjugate momentum $p$, which are evolved using classical equations of motion. The classical photonic trajectories are coupled to the evolution of the matter system via Ehrenfest forces; the matter may be treated quantum-mechanically which is usually the choice when only electronic degrees of freedom are explicitly considered, or also with mixed quantum-classical approaches where nuclear dynamics are also treated classically. A key aspect is the initial sampling of the photonic degrees of freedom: it has been shown that, starting in a zero-photon state, Wigner-sampling of the initial state is essential to capture the vacuum-fluctuations of the field and approximate photon-matter correlations well enough to describe, for example, the ensuing spontaneous emission of the matter system~\cite{HSRKA19,HSSRAK19,HLRM20}. 

This idea of Ref.~\cite{HSRKA19,HSSRAK19} to use a distribution of classical trajectories for the photonic system has enabled the investigation of new phenomena arising from strong light-matter coupling, including the importance of self-polarization when accounting for many modes in the cavity~\cite{HLRM20}, and modifications of chemical reaction rates from vibrational strong-coupling~\cite{LSN20,LMH21} where only the ground Born-Oppenheimer state is considered. Describing photons quasi-classically might seem, on the one hand, very surprising given that the photon is inherently a massless quantum particle. On the other hand, resolving the Hamiltonian in the photon displacement-coordinate representation renders the free-photon Hamiltonian a harmonic form, which, together with the bi-linear form of the light-matter coupling in the usual non-relativistic dipole-approximation treatments, makes it highly suitable for quasiclassical approximations. Indeed, classical Wigner propagation of the initial Wigner distribution is exact for quadratic Hamiltonians~\cite{H76}. However, even if the Hamiltonian has up to quadratic terms only in photonic displacement coordinate $q$,  coupling to the matter-system dynamics implicitly introduces a more complicated dependence on $q$ beyond quadratic, so we do not expect quasiclassical trajectory calculations to be exact despite the apparently quadratic form of the Hamiltonian. 

A closer look at the earlier studies indicate that the MTE approach, although quite accurate, tends to underestimate the photon number and intensities~\cite{HSRKA19, HSSRAK19, HLRM20}. To assess this, having an exactly numerically-solvable system to check against, as in these works,  is extremely valuable.
For example, in the studies of spontaneous emission from a two-level or a three-level system coupled to 400 cavity modes initially in vacuum, the intensity of the emitted light in the initial emission event is underestimated by about two-thirds (Figs. 3, 4, 7 and 8 of Ref.~\cite{HSRKA19}). A similar underestimation is seen in the photon number in the single-mode study of cavity-induced suppression of proton-coupled electron transfer in Fig 2d of Ref.~\cite{HLRM20}. 
A question arises: is this due to the quasiclassical approximation itself or from an inadequate accounting of light-matter correlation in the equations of motion driving the photons?
Quasiclassical approaches lack phases and interference effects, cannot describe classically-forbidden processes such as tunneling, and are liable not to fully hold onto zero-point energy.  The Wigner treatment in particular develops errors from strong anharmonicity. As mentioned earlier, although the terms in the light-matter Hamiltonian used in the computational models that involve the photonic displacement coordinate are purely harmonic with bi-linear coupling, the feedback with the matter system through this coupling effectively induces anharmonic effects on the photonic system. 

Whether it is the quasiclassical treatment itself that is largely responsible for the underestimation in the photon number and intensities, or whether it is instead largely due to incorrect forces in the classical equations driving the photon dynamics has not yet been determined. 
The latter reason was suggested in Ref.~\cite{HLRM20} by a qualitative argument  based on the exact factorization approach~\cite{AMG10,AMG12,HARM18,AG21}. In this approach exact time-dependent potentials are defined through exact time-dependent Schr\"odinger equations for coupled quantum subsystems. The original electron-nuclear formalism was extended to include photons in Ref.~\cite{HARM18}: depending on how the factorization is done, one can then obtain exact time-dependent potential energy surfaces (TDPES) for the nuclear dynamics~\cite{LHM19,MRHLM21}, for the electron dynamics~\cite{AKT18}, or for the photon dynamics~\cite{HARM18}.
 In the present work, we take a closer look at the exact potential driving the photon dynamics which we call the qTDPES. We show that the
underlying force in the MTE treatment of photons significantly differs from the exact, missing terms from  light-matter correlation. In particular,  the neglect of the photonic coordinate dependence of the coefficients that weigh the effective cavity-Born-Oppenheimer surfaces in evaluating the Ehrenfest forces has paramount importance. On the other hand, it is the quasiclassical approximation itself that leads to errors when the photon number per mode exceeds about a half. 

We begin in section~\ref{sec:model} by presenting the model we will study and its exact solution. 
In section~\ref{sec:methods} we introduce the particular flavor of the exact factorization approach used in this work, different approximate quasiclassical analyses of it and the MTE,  before presenting the results and analysis in Sec.~\ref{sec:results}. The implications for more realistic systems are discussed in Sec~\ref{sec:concs}, with prospects for effective practical methods based on the exact factorization approach. 

\section{Model system}
\label{sec:model}
The particular model we consider is simply one photon mode coupled resonantly to a two-level system, with Hamiltonian $H = H_m + H_p + H_{mp}$ where
\bea
\nonumber
H_m &=& \epsilon_g\ket{g}\bra{g} + \epsilon_e\ket{e}\bra{e}, \;
H_p = \left(\omega_c^2 q^2 - \partial^2/\partial q^2\right)/2\\
H_{mp} & = & r_{eg}\lambda \omega_c \,q \left( \ket{e}\bra{g} + \ket{g}\bra{e}\right)
\label{eq:H} 
\eea
Here $\ket{g (e)}$ are the two electronic states, while $q$ represents the displacement-field coordinate operator of the photonic system. 
Eq.~(\ref{eq:H}) follows from the non-relativistic Pauli-Fierz Hamiltonian~\cite{T13,RTFAR18,RFPATR14,CohenTannoudjiBook}, 
under the condition that the electronic states are parity eigenstates, and neglecting any dependence on nuclear variables. (No rotating wave approximation is made). The coupling matrix element $r_{eg} = \bra{e}\hat{r} \ket{g}$ where $\hat{r}$ is the (possibly many-)electron position operator. 
The Pauli-Fierz Hamiltonian also contains a self-polarization term, but in this simple two-level case, it can be absorbed into the matter part $H_m$, as a renormalization of the energies $\epsilon_g$ and $\epsilon_e$. We choose a cavity resonant with the electronic energy difference, $\omega_c = \epsilon_e  - \epsilon_g$. 

The exact solution may be expressed as the expansion
\ben
\ket{\Psi(q,t)} = \chi^d_g(q,t)\ket{g} + \chi^d_e(q,t)\ket{e} 
\label{eq:ex_diab}
\een
where $\vert\chi^d_{g (e)}\vert^2$ represent $q$-resolved populations associated with the uncoupled ground and excited electronic states respectively. The $\ket{ }$ notation refers to states in the electronic Hilbert space only. With analogy to the electron-nuclear problem, we may think of $\chi^d_{g (e)}(q,t)$ as the photonic state coefficients in a ``diabatic" expansion of the electronic system, since the electronic basis chosen in Eq.~(\ref{eq:ex_diab}) has no dependence on $q$. Alternatively, one can express the exact solution as
\ben
\ket{\Psi(q,t)} = \chi^{\rm qBO}_g(q,t)\ket{\Phi_{q,g}} + \chi^{\rm qBO}_e(q,t)\ket{\Phi_{q,e}}
\label{eq:ex_BO}
\een
where 
\ben
H^{\rm qBO}\ket{\Phi_{q,i}} ={\cal E}^{\rm qBO}_{i}(q)\ket{\Phi_{q,i}}
\label{eq:qBOdef}
\een
are cavity-Born-Oppenheimer states in the electronic Hilbert space, with
\ben
H^{\rm qBO} = H - T_q = H +\partial_q^2/2
\label{eq:HqBO}
\een
Eq.~(\ref{eq:ex_BO}) is analogous to the Born-Huang expansion in adiabatic states in the electron-nuclear case. 

The exact time-dependent solution is then given either by substituting Eq.~(\ref{eq:ex_diab}) or Eq.~(\ref{eq:ex_BO}) into the TDSE $H\ket{\Psi} = i\partial\ket{\Psi}{\partial t}$:
\bea
\nonumber
i\frac{\partial \chi^d_{g}(q,t)}{\partial t} &=& \left(-\partial_q^2/2 + \epsilon_{g} + \omega_c^2q^2/2\right)\chi^d_{g}(q,t) \\ &+& r_{eg}\lambda\omega_c \, q \chi^d_{e}(q,t)
\eea
and analogously for $\chi^d_e(q,t)$, 
or
\bea
\nonumber
i\frac{\partial \chi^{\rm qBO}_{g}(q,t)}{\partial t} &=& \left(-\partial_q^2/2 + {\cal E}^{\rm qBO}_{g}(q) - D_{gg}(q) \right)\chi^{\rm qBO}_{g}(q,t) \\ &-& \left(D_{ge}(q)  + d_{ge}(q)\partial_q\right)\chi^{\rm qBO}_e(q,t)
\eea
where the non-adiabatic coupling matrix elements are $D_{ij}(q)  = \bra{\Phi_{q,i}}\partial_q^2 \Phi_{q,j}\rangle/2$ and $d_{ge}(q)  = \bra{\Phi_{q,g}}\partial_q\Phi_{q,e}\rangle$.

\section{The MTE and exact factorization methods}
\label{sec:methods}
\subsection{Multi-Trajectory Ehrenfest for photons}
In the MTE approach, the displacement-coordinate of the photonic system is treated as a classical coordinate, and together with its conjugate variable $p$, define the phase-space for the classical trajectories. An ensemble of independent classical trajectories, initially sampled according to the initial photonic displacement density in $q$ and $p$, is propagated guided by a mean-field force determined by the electronic system, while the quantum nature of the electronic system is retained, driven by a potential parametrically dependent on the photon coordinate. Analogously to the Ehrenfest equations for the electron-nuclear problem, one obtains the equations for the $I$th photonic trajectory in the ensemble:
\bea
\label{eq:Ehrq}
\ddot{q}^{(I)} &=&  - \bra{\Phi^{(I)}(t)} \partial_q H^{\rm qBO} \ket{{\Phi^{(I)}}(t)}\\
i \partial_t \ket{\Phi^{(I)}(t)} &=& H_{\rm qBO}\ket{\Phi^{(I)}(t)},
\label{eq:Ehr}
\eea
where $\ket{\Phi^{(I)}(t)} = \ket{\Phi_{q^{(I)}}(t)}$, and the instantaneous classical position $q^{(I)}(t)$ replaces the photonic operator $q$ in the Hamiltonian $H_{\rm qBO}$. 
It is convenient to expand $\ket{\Phi^{(I)}(t)}$ in either the diabatic or Born-Huang basis, as in Eqs.~(\ref{eq:ex_diab})--(\ref{eq:ex_BO}) where the projected wavefunctions are now purely time-dependent coefficients associated with each trajectory, e.g. 
\ben
\ket{\Phi^{(I)}(t)} = \tilde c_g^{(I)}(t)\ket{g} + \tilde c_e^{(I)}(t)\ket{e} 
\een
for the diabatic case, replacing the superscript $d$ notation for this basis choice by a tilde, for convenience. Using the diabatic expansion, the equations become
\bea
\nonumber
\ddot{q}^{(I)}(t)   &=& -\omega_c^2 q^{(I)}(t)   - 2 \omega_c \lambda r_{eg} Re[\tilde c_g^{(I)*}(t)\tilde c_e^{(I)}(t)]\\
\dot{\tilde c}_{g (e)}^{(I)}(t)   &=& -i(\epsilon_g + \frac{\omega_c^2 q^{(I) 2}}{2}) \tilde c_{g (e)}^{(I)} - i\omega_c \lambda r_{eg} q^{(I)}  \tilde c_{e (g)}^{(I)}
\label{eq:Ehr_diab}
\eea
(where the $t$-dependence of $q^{(I)}, c_{g,(e)}$ is omitted on the right of the second equation to avoid notational clutter).
Instead, using the qBO basis, 
\ben
\ket{\Phi^{(I)}(t)} = c_g^{(I)}(t)\ket{\Phi_{q,g}} + c_e^{(I)}(t)\ket{\Phi_{q,e}} 
\een
we get 
\bea
\nonumber
\ddot{q}^{(I)}(t)   &=& -\sum_{i=g,e} \vert c^{(I)}_{i}\vert^2 \partial_q {\cal E}^{\rm qBO}_{i}
 -2Re[c^{(I)*}_{g}c^{(I)}_{e}] \Delta{\cal E}^{\rm qBO}d_{ge}\\
\dot{c}^{(I)}_{g (e)}(t) &=& - i {\cal E}^{\rm qBO}_{g (e)}c^{(I)}_{g (e)} - \dot{q}^{(I)}d_{eg}c^{(I)}_{e (g)}
\label{eq:Ehr_BO}
\eea
with $\Delta{\cal E}^{\rm qBO} = {\cal E}^{\rm qBO}_{e} - {\cal E}^{\rm qBO}_{g} $
(where the $q^{I}$-dependence of terms on the right is omitted for notational simplicity).

The Ehrenfest approach for the electron-nuclear problem was derived from treating nuclear degrees of freedom classically starting from a product ansatz of the molecular wavefunction~\cite{T98,MM80}; here we have applied it to the photonic system.  The potentials yielding the forces on the photonic displacement coordinate on the right-hand-side of Eq.~(\ref{eq:Ehrq}) have a mean-field character, seen perhaps most clearly in Eq.~(\ref{eq:Ehr_BO}) where the first two terms explicitly show the weighted BO-forces with an additional term coming from non-adiabatic transitions driven by the coupling $d_{eg}(q^{(I)})$. 

In the electron-nuclear case, the analogous mean-field force on the nuclear coordinate fails to be accurate after the system has propagated through
non-adiabatic interaction regions where the electronic eigenstates are strongly coupled. These regions tend to be somewhat localized in space, while we will see in Sec.~\ref{sec:results} (Fig.~\ref{fig:terms}) that in the present electron-photon case, the non-adiabatic coupling extends throughout the photonic displacement wavepacket. The Ehrenfest force is approximate at best, and instead, the exact factorization defines the exact force that drives the photonic system~\cite{HARM18,AMG10,AMG12}. 

\subsection{Exact forces from the exact factorization approach}
In the exact factorization approach, the exact time-dependent wavefunction of a system of coupled quantum subsystems is written as a single correlated product of marginal and conditional factors: in the present case,
\ben
\ket{\Psi(q,t)} =  \chi(q,t)\ket{\Phi_q(t)},\;{\rm with} \; \langle\Phi_q(t)\ket{\Phi_q(t)} = 1
\label{eq:XF}
\een
where the marginal $\chi(q,t)$ represents the photonic wavefunction and $\ket{\Phi_q(t)}$ is the electronic state conditionally dependent on the photonic displacement-field coordinate $q$. The equality on the right is referred to as the ``partial normalization condition", and indicates that  the conditional wavefunction is normalized to 1 for every $q$ at each time $t$; the notation $\langle ...|  ...\rangle$ means an inner-product only over the electronic Hilbert space. Despite being a single product (in contrast to Eq.~(\ref{eq:ex_BO})), Eq.~(\ref{eq:XF}) represents the exact wavefunction, and the two factors are unique up to a $(q,t)$-dependent phase factor: $e^{i\theta(q,t)}$ can multiply $\chi(q,t)$ while $e^{-i\theta(q,t)}$  multiplies $\ket{\Phi_q(t)}$ to yield the same $\ket{\Psi(q,t)} = \chi(q,t)\ket{\Phi_q(t)}$. The equations for the two components involve terms that depend on the other factor, which intimately represent the electron-photon correlation; these can be found in Ref.~\cite{HARM18} and are straightforward generalizations of the electron-nuclear case to the electron-photon one. In particular, with the present choice of factorization, the equation for $\chi(q,t)$ has a time-dependent Schr\"odinger form with a vector potential and a scalar potential.

The phase-freedom appears as a gauge-like freedom in the electron-nuclear coupling potentials, and, in some cases, a gauge can be selected such that only a scalar potential appears. This is always the case when the marginal has only one degree of freedom. (We note that in the general case of $N$ photon modes in the cavity, where the marginal is $N$-dimensional, the vector potential is also $N$-dimensional, and has the form $A_\nu = \langle \Phi_{\bq}(t) \ket{i\partial_{q_\nu} \Phi_\bq(t)}$. 
Even though each mode may be one-dimensional, the vector potential may still have a rotational component. ) With only one mode in our model system, we may choose a gauge where $A = 0$, and then we find that
the potential in the Schr\"odinger equation for $\chi(q,t)$ is purely scalar and time-dependent, and we denote it the qTDPES:
\ben
{\cal E}^{\rm qTDPES}(q,t) = {\cal E}^{\rm wBO}(q,t)  + {\cal E}^{\rm kin}(q,t) + {\cal E}^{\rm GD}(q,t)
\label{eq:qtdpes}
\een
where 
\ben
{\cal E}^{\rm wBO}(q,t) = \bra{\Phi_q(t)}H^{\rm qBO}\ket{\Phi_q(t)}
\label{eq:wBO}
\een
\ben
{\cal E}^{\rm kin}(q,t) =\bra{\partial_q \Phi_q(t)}\partial_q \Phi_q(t)\rangle/2
\label{eq:ekin}
\een
and
\ben
{\cal E}^{\rm GD}(q,t) = \bra{\Phi_q(t)}-i\partial_t\Phi_q(t)\rangle
\label{eq:eGD}
\een

This allows us to define an exact force driving the photons in an exact-classical-trajectory approach, via
\ben
\ddot{q}^{(I)}(t) = -\partial_q {\cal E}^{\rm qTDPES}(q^{(I)},t)
\label{eq:XFforce}
\een

A question we will ask, is how well do the classical trajectories for the photons evolving under the exact force, Eq.~(\ref{eq:XFforce}), do, compared with the Ehrenfest photon trajectories, Eq.~(\ref{eq:Ehr_BO}) or Eq.~(\ref{eq:Ehr_diab}), in reproducing the true photonic dynamics.

To get more insight into the exact force, it is instructive to expand the conditional electronic wavefunction in terms of the qBO wavefunctions that were defined in Eq.~(\ref{eq:qBOdef}):
\ben
\ket{\Phi_q(t)} = C_g(q,t)\ket{\Phi_{q,g}} + C_e(q,t)\ket{\Phi_{q,e}} 
\een
where, due to the partial normalization condition, $\vert C_g(q,t)\vert^2 + \vert C_e(q,t)\vert^2 =1$ for every $q$ and $t$. 
In terms of these coefficients, we observe that the force from the weighted BO contribution to ${\cal E}^{\rm qTDPES}(q^{(I)},t)$ somewhat resembles the force in the MTE:
\bea
\nonumber
\label{eq:wBOforce1}
\ddot{q}^{(I), {\rm wBO}}(t)&=& -\partial_q \left(\vert C_{g}\vert^2 {\cal E}^{\rm qBO}_{g}+ \vert C_{e}\vert^2 {\cal E}^{\rm qBO}_{e}\right)\\
&=& -\sum_{i=g,e} \left(\vert C_{i}\vert^2 \partial_q {\cal E}^{\rm qBO}_{i} - (\partial_q \vert C_i\vert^2){\cal E}^{\rm qBO}_{i} \right)\\
\nonumber
& =& -\partial_q{\cal E}^{\rm qBO}_{g} - \vert C_e\vert^2\partial_q({\cal E}^{\rm qBO}_{e} - {\cal E}^{\rm qBO}_{g} ) \\
&-& (\partial_q \vert C_e\vert^2)({\cal E}^{\rm qBO}_{e} - {\cal E}^{\rm qBO}_{g})
\label{eq:wBOforce2}
\eea
We will see that the two qBO surfaces are almost parallel, both very close to harmonic with deviations only at larger $q$ while the coefficients show a large $q$-dependence near $q=0$ (see Sec.~\ref{sec:results}): as a result, the wBO force is dominated by an almost linear restoring force (first term) together with a large repulsive force away from $q=0$ from the last term in Eq.~(\ref{eq:wBOforce2}).

Comparing now with Eq.~(\ref{eq:Ehr_BO}) we see that the Ehrenfest force has the similar structure of a weighted force from the first term on the right of Eq.~(\ref{eq:wBOforce1}), but a crucial difference in the last term: while the local shape of the coefficients plays a large role in the wBO force, this is completely absent in the Ehrenfest force, which instead has a term depending on the non-adiabatic coupling between the two electronic states. We will find that this term is much smaller than that from the $q$-dependence of the coefficients appearing in the wBO term and, in contrast to the large $q$-dependence of this term, this term in the MTE gives an almost uniform force in $q$. The Ehrenfest method involves independent trajectories and the notion of how a coefficient associated with one trajectory varies with that of a neighboring trajectory does not arise with these uncoupled trajectories.

\subsection{Observables from the exact factorization approach}
The main observable of interest in this study is the photon number, 
$\langle N\rangle$, defined by 
\ben
\langle N\rangle = \langle a^\dagger a \rangle = \langle \omega q^2 + p^2/\omega \rangle/2 - 1/2
\een
(As mentioned earlier, there is no contribution from self-polarization in this model).  
The distribution of photonic trajectories in the $q,p$ phase space in the MTE calculation directly yields the expectation values on the right from the its variances in position and momentum. When extracting these from the exact factorization framework, a subtlety arises: the marginal wavefunction of the factorization, $\chi(q,t)$, reproduces the photonic density and photonic current-density as would be obtained from the full photon-electron wavefunction, $\ket{\Psi(q,t)}$, but for general non-multiplicative observables of the photonic system, the $q$-dependence of the conditional wavefunction means that there are additional terms to the usual operators. Specifically~\cite{AMG12}, for our choice of gauge,
\bea
\nonumber
\langle p^2 \rangle &=& \int dq \bra{\Psi(t)} -\partial_q^2 \Psi(t)\rangle \\
\nonumber
&=& -\int dq\left(\chi^*(q,t)\partial_q^2\chi(q,t) - \vert\chi(q,t)\vert^2\bra{\Phi_q(t)} \partial_q^2 \Phi_q(t)\rangle\right)\\ 
& =& \langle p^2 \rangle_\chi + 2\int dq \vert \chi(q,t)\vert^2{\cal E}_{\rm kin}(q,t)
\label{eq:psq-calc}
\eea
where $\langle p^2 \rangle_\chi  = -\int dq \chi^*(q,t)\partial_q^2\chi(q,t)$ is twice the kinetic energy of the marginal. The second term in Eq.~(\ref{eq:psq-calc}) must be added to this in order to get the true expectation value of $p^2$. 
This applies both for cases where a quantum calculation is made for the marginal $\chi(q,t)$, as well as when it is computed through a quasiclassical trajectory distribution in $(q,p)$-phase space; in the latter case the $q$-density weighted average of ${\cal E}_{\rm kin}$ must be added to the variance in $p$. 

\section{Results}
\label{sec:results}
We will now analyze the results of the MTE approach for photons by comparing against quantum and quasiclassical propagation on the exact qTDPES, as well as from propagation on only the wBO component. Due to the simplicity of the model system of Sec.~\ref{sec:model},  the numerically exact solution for $\ket{\Psi(q,t)}$ is easily obtained. 
The exact qTDPES and the wBO component are found by inversion: writing $\chi(q,t) = \vert\chi(q,t)\vert e^{iS(q,t)}$, where $\vert\chi(q,t)\vert = \sqrt{\bra{\Psi(q,t)}\Psi(q,t)\rangle}$, and $\partial_q S(q,t) = {\rm Im}\bra{\Psi(q,t)}\partial_q\Psi(q,t)\rangle/\vert\chi(q,t)\vert^2$, we can find $\ket{\Phi_q(t)} = \ket{\Psi(q,t)}/\chi(q,t)$ to insert into Eqs.~(\ref{eq:qtdpes})--(\ref{eq:eGD}). The expression for the phase $S(q,t)$ follows from the gauge choice $A = 0$, as in Refs.~\cite{AASMMG15,HARM18}.

We choose parameters for the model of Eq.~(\ref{eq:H}) as $\epsilon_e = 0.2, \epsilon_g = -0.2, \omega_c = 0.4$a.u. and $r_{eg}\lambda = 0.01$a.u. The resulting dynamics gives Rabi-like oscillations, with a period of $T\approx 2\pi/0.01$. 

Before we examine the results of the quasiclassical propagations, we first compare the underlying potentials and forces in MTE dynamics with those arising from the exact factorization. 

\subsection{Forces from MTE in light of exact factorization}
\label{sec:forces}
\begin{figure}[h]
\includegraphics[width=0.5\textwidth]{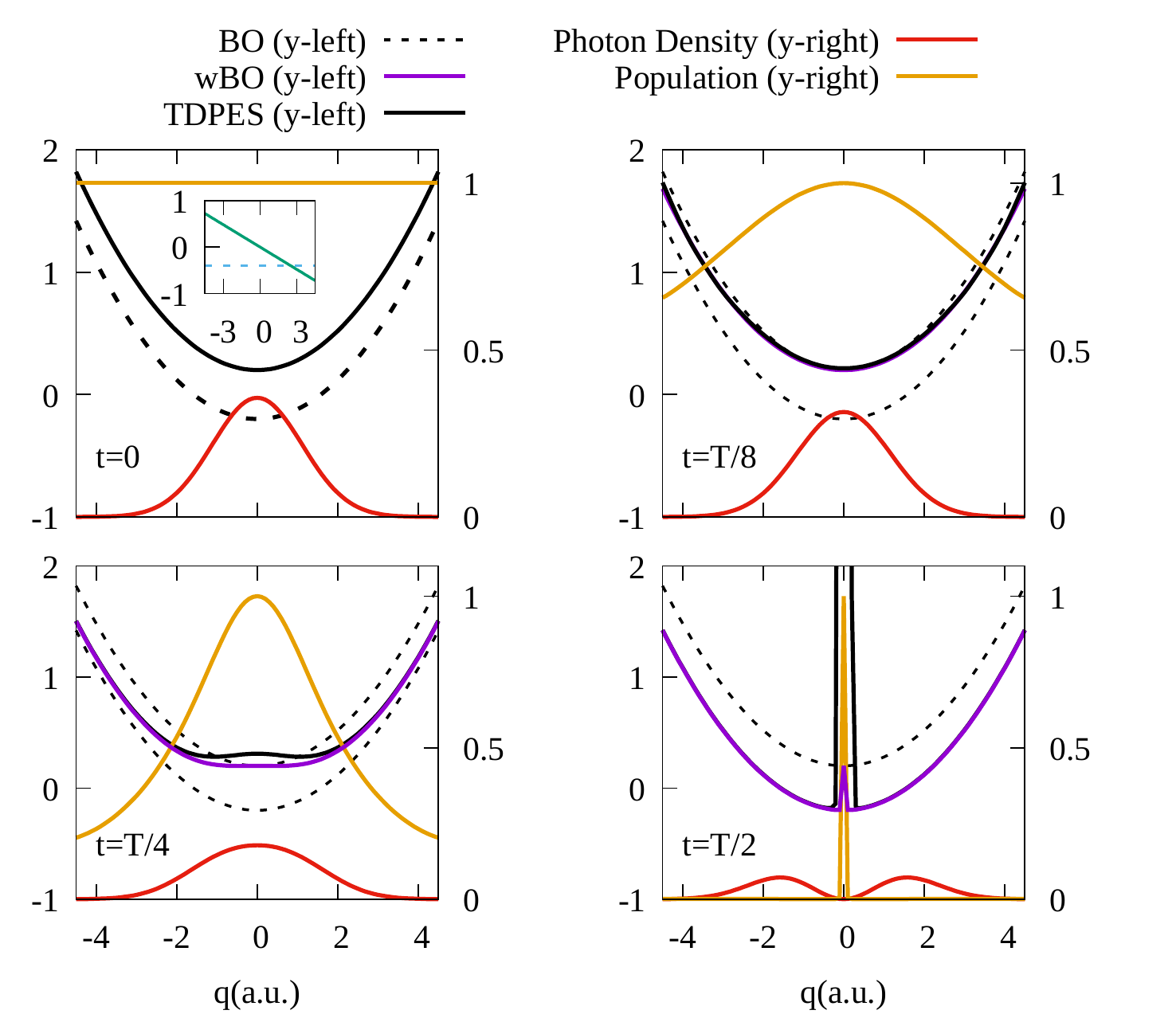}
\caption{Time-snapshots over half a Rabi-cycle, showing the exact qTDPES (black solid), the wBO surface (purple), the qBO surfaces (black dashed), the photonic density $\vert\chi(q,t)\vert^2$ (red), and the population of the upper electronic level $\vert C_e(q,t)\vert^2$ (orange); the $y$-scale for the latter two is on the right. The inset in the $t=0$ plot  shows the terms determining the two force contributions in the MTE Eq.~(\ref{eq:Ehr_BO}), $\partial_q{\cal E}^{\rm qBO}_g \approx \partial_q{\cal E}^{\rm qBO}_e$ (green solid) and $d_{eg}(q)({\cal E}^{\rm qBO}_e(q)-{\cal E}^{\rm qBO}_e(q))$ (blue dashed, scaled by 100). }
\label{fig:terms}
\end{figure}

The top left panel of Fig.~\ref{fig:terms} shows the two qBO surfaces, which have a form that deviates from harmonic by less than 0.01\% in the range shown: diagonalization yields ${\cal E}^{\rm qBO}(q) = \omega_c^2 q^2/2 \mp \sqrt{\omega_c^2/4 + (r_{eg}\lambda\omega_c)^2q^2}$, which shows that the deviation grows with $q^2$.
The plot also shows the non-adiabatic coupling $d_{eg}(q)$, indicating that the MTE force in Eq.~(\ref{eq:Ehr_BO}) is dominated by the first term, and also that, unlike typical electron-nuclear case, the non-adiabatic coupling in the electron-photon case is delocalized in $q$-space.
The force acting on the MTE trajectories can be compared with that acting on trajectories evolving on the exact qTDPES and wBO surfaces, and to discuss these we must discuss the initial state since these surfaces are both time-dependent and state-dependent. 

We begin with the system initially in the state $\ket{\Psi(q,0)} = \chi_g(q)\ket{\Phi_{q,e}}$ where the  photon ground state is the zero-photon state $\chi_g(q) = \left( \frac{\omega}{\pi}\right)^{1/4}\exp(-\omega q^2/2)$. 
Because the conditional electronic state is initially in the excited eigenstate of $H_{\rm qBO}$, the exact qTDPES and wBO surfaces coincide with the upper qBO surface at the initial time, but quickly begin to deviate from it. After short times, anharmonicity develops  at larger $q$ in the wBO and qTDPES, which begin to move towards the lower qBO surface, while at smaller $q$, the surfaces remain close to the upper qBO surface. This is consistent with the bi-linear nature of the electron-photon coupling scaling with $q$, and we see that the $q$-resolved electronic population of the upper qBO surface, shown in the figures, pulls away from the excited state to the lower at larger $q$. Over time, a barrier builds up near $q=0$, at which point the wBO resolutely sticks to the upper qBO surfaces where the electron-photon coupling is locally zero, while everywhere else collapsing to the lower qBO surface, associated with the photon emission. The inclusion of the kinetic and gauge-dependent terms to form the full qTDPES, lead to an even more pronounced and increasingly localized barrier at $q = 0$, arising primarily from ${\cal E}_{\rm kin}(q,t)$~\cite{HARM18}, which becomes singular at $T/2$. During this half-cycle, the photon emission character is evident in $\vert\chi(q,t)\vert^2$ which evolves from the harmonic ground-state form to the first-excited state (one-photon state), and also in the population of the excited state $|C_e(q,t)|^2$ which evolves from  initially being $1$ at all $q$ to $0$ at all $q$ except for $q=0$. 

The correct forces in a quasiclassical description of the photons are thus significantly different to what is in the MTE approach. 
Clearly the anharmonicity and developing barrier in the wBO and qTDPES surfaces  will yield different forces to that of the MTE, leading to a broader distribution in $q$, which implies greater $\langle q^2 \rangle$ and greater photon number. Relating this back to Eq.~(\ref{eq:wBOforce2}), where the wBO force is decomposed in terms of qBO components and coefficients, we had noted that the first term is shared with the MTE, the second term is negligible (also clear from the figure), while the third term gives rise to the anharmonicity and barrier: the $q$-dependence of the electronic coefficient, which intimately reflects the electron-photon correlation, is intimately is missing in MTE.

The lack of this electron-photon correlation barrier and large-$q$ widening in the potential underlying MTE, is expected to lead to an underestimation in photon number. 
A separate question is then how well does the quasiclassical treatment itself, in contrast to a quantum treatment, perform for dynamics with such anharmonic structures? Even if the exact factorization provides the correct forces that a classical particle would experience, if the classical approximation for the trajectory ensemble is itself poor, it is not of practical use. In Sec.~\ref{sec:prop} we turn to the propagation itself.

\subsection{Quasiclassical and quantum propagations: MTE, wBO, and qTDPES}
\label{sec:prop}
We now verify the predicted underestimation by running quasiclassical dynamics on the exact factorization surface qTDPES, and  the wBO component, and comparing these with MTE. It is also instructive to run a quantum dynamics on the qTDPES and the wBO component: we will find that as the photon emission approaches about half a photon and proceeds beyond that, the quasiclassical approximation itself  becomes inaccurate, due to 
the growing anharmonicity of the potential that includes the increasingly narrow and sharp barrier near $q = 0$. 

The quantum dynamics is computed with the split-operator method with a time-step of $0.001$a.u. and a spatial grid-spacing of 0.1a.u. 
The quantum dynamics run on the full qTDPES agreed with the results from the original full molecular propagation, which served  as a test that our inversion to find the exact qTDPES was accurate. We refer to this below as the exact quantum dynamics. 
The quasiclassical propagation was carried out by solving Hamilton's equations with a time-step of 0.02 a.u. via leapfrog integration with 10,000 trajectories (initially sampled using the Gaussian initial state).

\begin{figure}[h]
\includegraphics[width=0.5\textwidth]{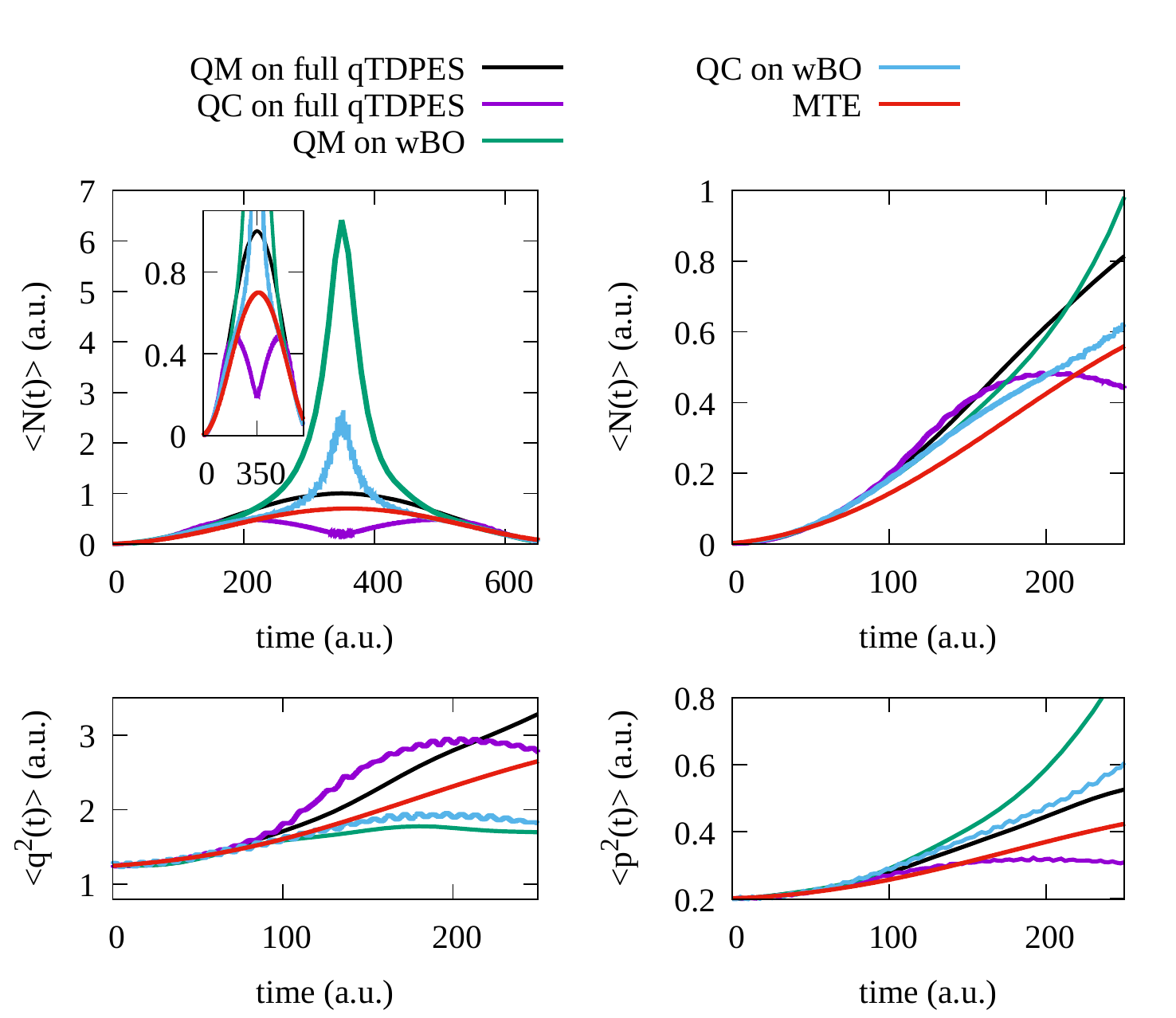}
\caption{Top row: the photon number $\langle N(t)\rangle$ as computed through the indicated methods. Lower row shows the field intensities $\langle q^2(t)\rangle$ and  $\langle p^2(t)\rangle$.}
\label{fig:averages}
\end{figure}

\begin{widetext}
\begin{figure*}[ht]
\includegraphics[width=0.8\textwidth,angle=270]{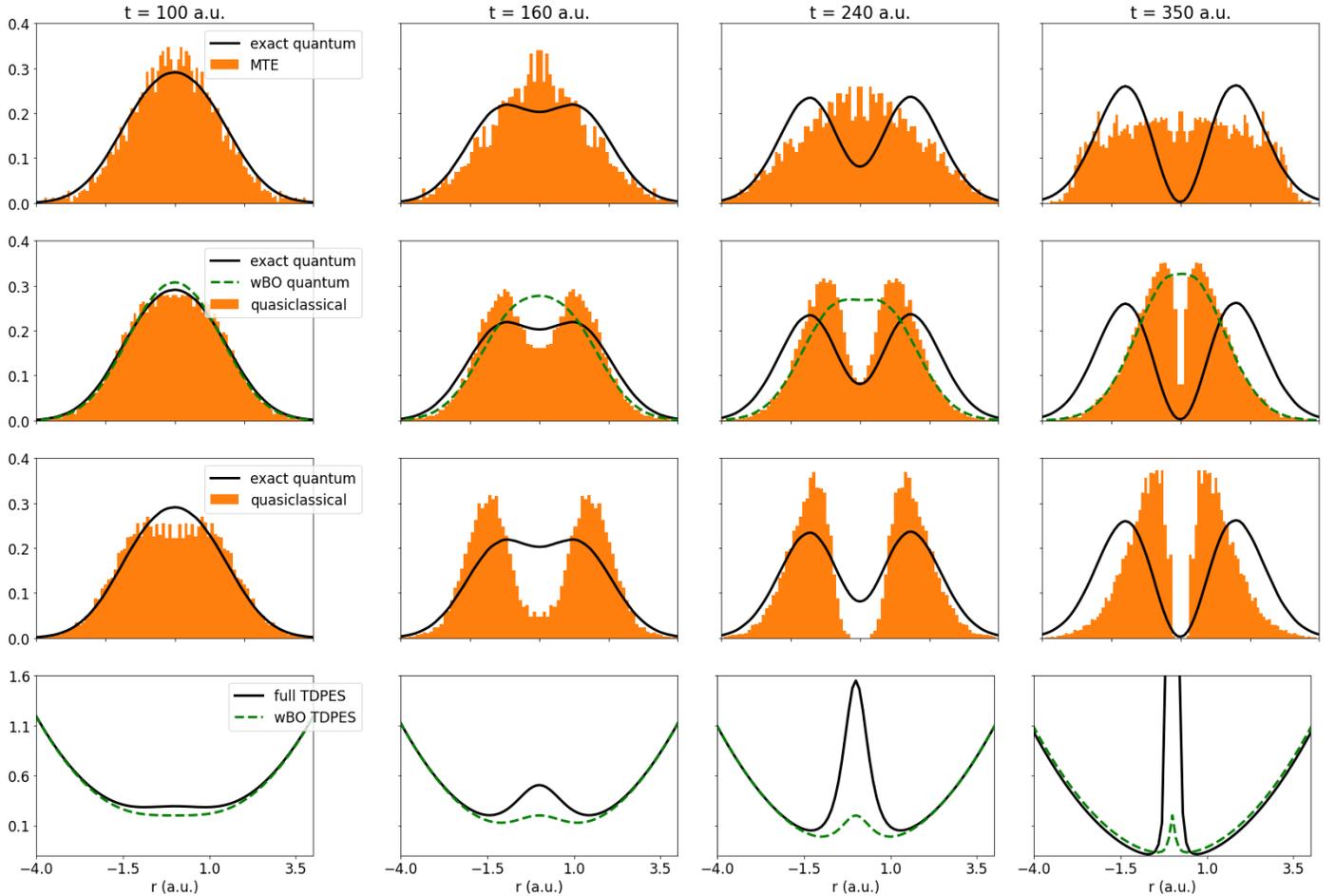}
\caption{Time snapshots of the photonic displacement coordinate density, computed via MTE (top), quasiclassical (orange) and quantum (green dashed) on the wBO surface along with the exact quantum (black) for reference (second row), and the quasiclassical (orange) and quantum (black) on the full qTDPES (third row), shown at a) $t= 100$au, b) $t = 160$au, c) $t = 240$au, d) $t= 350$au. The fourth  row shows the exact qTDPES (black) along with the wBO component (green dashed). }
\label{fig:snaps}
\end{figure*}
\end{widetext}

The top left panel and inset of Fig.~\ref{fig:averages} shows the photon number during a full Rabi period, computed from the exact quantum dynamics which shows the expected emission of one photon after half a Rabi period. The quantum dynamics on the wBO component appears to be reasonably accurate until about half a photon is emitted, with a slight underestimate, which would be expected from the smaller barrier at $q=0$ observed in the potential in Fig.~\ref{fig:terms}. However after about a quarter Rabi period the quantum propagation on the wBO begins to greatly overestimate $\langle N \rangle$, largely from the ${\cal E}_{\rm kin}$ contribution to $\langle p^2 \rangle$: the peak in this term at $q=0$ grows increasingly sharp and localized as the one-photon state is reached, and because the photonic density remains significant in this region due to the lack of this term in determining the trajectory evolution, it samples the peak and leads to large values. This is also clear from Fig.~\ref{fig:snaps}, where in the second row, the green dashed  curve shows the photon displacement-coordinate density obtained from quantum propagation on the wBO surface, and the fourth row shows  the wBO surface itself as the green dashed curve, at four time-snapshots. The small barrier at $q=0$ can be tunneled through and so leads to only a small dip in the density there at about a quarter-Rabi period.  Whether this problem appears in a self-consistent calculation, where the terms in the qTDPES are computed by a self-consistent conditional electronic state rather than the exact one,  is a question for future investigation. 

Turning now to the quasiclassical propagations, we see from Fig.~\ref{fig:averages} that for the first quarter Rabi period or so, quasiclassical propagation on both the full qTDPES and wBO are reasonably accurate,  especially for the photon number where there is some error cancellation between a small overestimate of $\langle q^2(t)\rangle$ and underestimate of $\langle p^2(t)\rangle$. The MTE underestimates the photon number $\langle N(t)\rangle$ and the intensities $\langle q^2(t)\rangle$, $\langle p^2(t)\rangle$. 
The time snapshots in Fig.~\ref{fig:snaps} enable a deeper analysis and movies are also provided in the Supplementary Information. 

Let us first discuss the MTE case. The top figures in each of the four panels in Fig.~\ref{fig:snaps} shows the MTE photonic displacement-coordinate density, which shows a ``breathing" that leads to the larger width, but remain at all times peaked around $q=0$ in stark contrast to the exact photonic density. The anharmonicity at large $q$ in the first term of Eq.~\ref{eq:Ehr_BO} along with the non-adiabatic coupling cause the density to spread, somewhat similar to a harmonic oscillator that widens (and narrows in the second half of the cycle). But without any barrier term,  MTE is unable to even remotely capture the correct structure of the density. The increase in the averaged quantities such as $\langle q^2\rangle$ and $N$ as a photon is emitted give an overly rosy picture of the photonic dynamics, which likely lead to errors in other properties of the radiation field. 
The  MTE underestimation was expected from the discussion of Sec.~\ref{sec:forces} and here we see it explicitly in the dynamics.  For the photon number, it is of similar magnitude seen in the previous works on a model molecule and with many modes, as discussed in Sec.~\ref{sec:intro} for the specific examples where exact results were available for comparison.  

We now consider the quasiclassical dynamics on the exact qTDPES.  Looking at the time-snapshots in Fig.~\ref{fig:snaps} in the third row, we notice that the density splits too quickly in the early dynamics ($t = 100, 160$au in the figure). This may be explained by the fact that
the trajectories are pushed away from the potential barrier developing
in the middle, while the quantum wavepacket can still live in the classically
forbidden region. The quasiclassical underestimates the outer tail regions of
the density slightly but overestimates the peaks in order
to compensate the density that has been shunned away from the
barrier. When the potential barrier  becomes larger, we observe that the classical
trajectories tend to over-localize in each (partial) well. 
The two errors somewhat compensate in the averaged quantities --  less density in the classically-forbidden region near $q = 0$, but also less density in the tails -- leading to the reasonably accurate averages observed earlier in Fig.~\ref{fig:averages}.

The average $\langle N(t)\rangle$ from the quasiclassical propagation on the wBO surface appears to performs about the same  for much of the quarter Rabi period, as that on the full qTDPES, due to an error cancellation effect: the underlying potential displays a smaller barrier at the origin so the trajectories are not repelled away as much. In fact the quasiclassical $q$-resolved displacement density is more accurate  for propagation on wBO than for propagation on the full qTDPES, as evident from the second row in Fig.~\ref{fig:snaps}. 
At larger times, there is an underestimate in $\langle q^2\rangle$ which is compensated by an overestimate of $\langle p^2\rangle$ which comes from the ${\cal E}_{\rm kin}$ peak in Eq.~(\ref{eq:psq-calc}) which is sampled more because the photonic density does not split as much. 


\section{Conclusions and Outlook}
\label{sec:concs}
Although the present work considered only the very simplest case of a two-level system coupled to a single photon-mode, we expect that the essential results generalize to realistic systems with many photon-modes, many electronic levels, and nuclear degrees of freedom: an MTE treatment of photons tends to underestimate the photon number and intensities, due to an inadequate accounting of matter-photon coupling in the treatment of the independent-trajectory Ehrenfest coefficients. The underlying photonic displacement-coordinate density does not show the character expected from the partial emission of a photon. The underestimation observed in the simple model has about the same magnitude as that observed in more complex systems where exact results were available for comparison~\cite{HSRKA19,HSSRAK19,HLRM20}. 

In contrast, a quasiclassical evolution on the weighted qBO surfaces, wBO, or full qTDPES performs much better for photon numbers of less than a half per mode,
 while for larger photon numbers per mode, large errors arise from the quasiclassical approximation itself that performs poorly when the underlying potential changes rapidly in space and becomes far from locally quadratic. The quasiclassical calculations were performed here within a particular gauge choice for the exact factorization (zero vector potential), and an open question is whether a different gauge choice would result in more accurate results at longer times. The sharp barrier forming at $q=0$ arises in the gauge-independent term ${\cal E}_{\rm kin}$ however while in this case the gauge-dependent term is relatively small~\cite{HARM18}. 
 
Still, the computational efficiency and reasonable accuracy of MTE treatment of photonic dynamics in cavity-QED is encouraging~\cite{HSRKA19,HSSRAK19,HLRM20,LSN20}, and the present results should be taken more as an explanation of its errors and as a note of caution that in such calculations the photon numbers are likely underestimated. Global quantities obtained from averaging over the Ehrenfest trajectories may appear more accurate than the underlying photonic displacement-coordinate density, which likely has a qualitatively incorrect structure.
(We also note that  with MTE methods, zero point energy leakage from higher modes to lower modes may occur, and should be checked; this was not a problem with the single-mode calculation here). 

Again, it is important to note that the quasiclassical calculations on the exact-factorization surfaces are not self-consistent ones, {\it i.e.} we are using the surfaces extracted from inversion of an exact calculation and simply running quasiclassical dynamics on them. We are using them as an analysis tool rather than as a method in itself. A self-consistent mixed quantum-classical calculation based on the exact factorization  will be investigated in the future, for example extending those developed for the electron-nuclear problem~\cite{MAG15,AMAG16,MATG17,AG21,HLM18,VIHMCM21}.

\begin{acknowledgments}{Financial support from the US National Science Foundation
CHE-1940333 (N.T.M and B.R. ) and the Department of Energy, Office
of Basic Energy Sciences, Division of Chemical Sciences,
Geosciences and Biosciences under Award DE-SC0020044(L.L.)
are gratefully acknowledged. The work of N.M.H was supported by the German National Academy of Sciences Leopoldina. }
\end{acknowledgments}

\bibliography{./ref}

\begin{thebibliography}{47}%
\makeatletter
\providecommand \@ifxundefined [1]{%
 \@ifx{#1\undefined}
}%
\providecommand \@ifnum [1]{%
 \ifnum #1\expandafter \@firstoftwo
 \else \expandafter \@secondoftwo
 \fi
}%
\providecommand \@ifx [1]{%
 \ifx #1\expandafter \@firstoftwo
 \else \expandafter \@secondoftwo
 \fi
}%
\providecommand \natexlab [1]{#1}%
\providecommand \enquote  [1]{``#1''}%
\providecommand \bibnamefont  [1]{#1}%
\providecommand \bibfnamefont [1]{#1}%
\providecommand \citenamefont [1]{#1}%
\providecommand \href@noop [0]{\@secondoftwo}%
\providecommand \href [0]{\begingroup \@sanitize@url \@href}%
\providecommand \@href[1]{\@@startlink{#1}\@@href}%
\providecommand \@@href[1]{\endgroup#1\@@endlink}%
\providecommand \@sanitize@url [0]{\catcode `\\12\catcode `\$12\catcode
  `\&12\catcode `\#12\catcode `\^12\catcode `\_12\catcode `\%12\relax}%
\providecommand \@@startlink[1]{}%
\providecommand \@@endlink[0]{}%
\providecommand \url  [0]{\begingroup\@sanitize@url \@url }%
\providecommand \@url [1]{\endgroup\@href {#1}{\urlprefix }}%
\providecommand \urlprefix  [0]{URL }%
\providecommand \Eprint [0]{\href }%
\providecommand \doibase [0]{http://dx.doi.org/}%
\providecommand \selectlanguage [0]{\@gobble}%
\providecommand \bibinfo  [0]{\@secondoftwo}%
\providecommand \bibfield  [0]{\@secondoftwo}%
\providecommand \translation [1]{[#1]}%
\providecommand \BibitemOpen [0]{}%
\providecommand \bibitemStop [0]{}%
\providecommand \bibitemNoStop [0]{.\EOS\space}%
\providecommand \EOS [0]{\spacefactor3000\relax}%
\providecommand \BibitemShut  [1]{\csname bibitem#1\endcsname}%
\let\auto@bib@innerbib\@empty
\bibitem [{\citenamefont {Ebbesen}(2016)}]{E16}%
  \BibitemOpen
  \bibfield  {author} {\bibinfo {author} {\bibfnamefont {T.~W.}\ \bibnamefont
  {Ebbesen}},\ }\href@noop {} {\bibfield  {journal} {\bibinfo  {journal}
  {Accounts of Chemical Research}\ }\textbf {\bibinfo {volume} {49}},\ \bibinfo
  {pages} {2403} (\bibinfo {year} {2016})}\BibitemShut {NoStop}%
\bibitem [{\citenamefont {Herrera}\ and\ \citenamefont
  {Owrutsky}(2020)}]{HO20}%
  \BibitemOpen
  \bibfield  {author} {\bibinfo {author} {\bibfnamefont {F.}~\bibnamefont
  {Herrera}}\ and\ \bibinfo {author} {\bibfnamefont {J.}~\bibnamefont
  {Owrutsky}},\ }\href@noop {} {\bibfield  {journal} {\bibinfo  {journal} {The
  Journal of Chemical Physics}\ }\textbf {\bibinfo {volume} {152}},\ \bibinfo
  {pages} {100902} (\bibinfo {year} {2020})}\BibitemShut {NoStop}%
\bibitem [{\citenamefont {Garcia-Vidal}\ \emph {et~al.}(2021)\citenamefont
  {Garcia-Vidal}, \citenamefont {Ciuti},\ and\ \citenamefont
  {Ebbesen}}]{GCE21}%
  \BibitemOpen
  \bibfield  {author} {\bibinfo {author} {\bibfnamefont {F.~J.}\ \bibnamefont
  {Garcia-Vidal}}, \bibinfo {author} {\bibfnamefont {C.}~\bibnamefont {Ciuti}},
  \ and\ \bibinfo {author} {\bibfnamefont {T.~W.}\ \bibnamefont {Ebbesen}},\
  }\href@noop {} {\bibfield  {journal} {\bibinfo  {journal} {Science}\ }\textbf
  {\bibinfo {volume} {373}},\ \bibinfo {pages} {eabd0336} (\bibinfo {year}
  {2021})}\BibitemShut {NoStop}%
\bibitem [{\citenamefont {Hertzog}\ \emph {et~al.}(2019)\citenamefont
  {Hertzog}, \citenamefont {Wang}, \citenamefont {Mony},\ and\ \citenamefont
  {B{\"o}rjesson}}]{HWMB19}%
  \BibitemOpen
  \bibfield  {author} {\bibinfo {author} {\bibfnamefont {M.}~\bibnamefont
  {Hertzog}}, \bibinfo {author} {\bibfnamefont {M.}~\bibnamefont {Wang}},
  \bibinfo {author} {\bibfnamefont {J.}~\bibnamefont {Mony}}, \ and\ \bibinfo
  {author} {\bibfnamefont {K.}~\bibnamefont {B{\"o}rjesson}},\ }\href@noop {}
  {\bibfield  {journal} {\bibinfo  {journal} {Chemical Society Reviews}\
  }\textbf {\bibinfo {volume} {48}},\ \bibinfo {pages} {937} (\bibinfo {year}
  {2019})}\BibitemShut {NoStop}%
\bibitem [{\citenamefont {Feist}(2021)}]{F21}%
  \BibitemOpen
  \bibfield  {author} {\bibinfo {author} {\bibfnamefont {J.}~\bibnamefont
  {Feist}},\ }\href@noop {} {\bibfield  {journal} {\bibinfo  {journal}
  {Nature}\ }\textbf {\bibinfo {volume} {597}},\ \bibinfo {pages} {185}
  (\bibinfo {year} {2021})}\BibitemShut {NoStop}%
\bibitem [{\citenamefont {Munkhbat}\ \emph {et~al.}(2021)\citenamefont
  {Munkhbat}, \citenamefont {Canales}, \citenamefont {K{\"u}c{\"u}k{\"o}z},
  \citenamefont {Baranov},\ and\ \citenamefont {Shegai}}]{Munkhbat2021}%
  \BibitemOpen
  \bibfield  {author} {\bibinfo {author} {\bibfnamefont {B.}~\bibnamefont
  {Munkhbat}}, \bibinfo {author} {\bibfnamefont {A.}~\bibnamefont {Canales}},
  \bibinfo {author} {\bibfnamefont {B.}~\bibnamefont {K{\"u}c{\"u}k{\"o}z}},
  \bibinfo {author} {\bibfnamefont {D.~G.}\ \bibnamefont {Baranov}}, \ and\
  \bibinfo {author} {\bibfnamefont {T.~O.}\ \bibnamefont {Shegai}},\ }\href
  {\doibase 10.1038/s41586-021-03826-3} {\bibfield  {journal} {\bibinfo
  {journal} {Nature}\ }\textbf {\bibinfo {volume} {597}},\ \bibinfo {pages}
  {214} (\bibinfo {year} {2021})}\BibitemShut {NoStop}%
\bibitem [{\citenamefont {Esteso}\ \emph {et~al.}(2021)\citenamefont {Esteso},
  \citenamefont {Cali{\`o}}, \citenamefont {Espin{\'o}s}, \citenamefont
  {Lavarda}, \citenamefont {Torres}, \citenamefont {Feist}, \citenamefont
  {Garc{\'\i}a-Vidal}, \citenamefont {Bottari},\ and\ \citenamefont
  {M{\'\i}guez}}]{Esteso2021}%
  \BibitemOpen
  \bibfield  {author} {\bibinfo {author} {\bibfnamefont {V.}~\bibnamefont
  {Esteso}}, \bibinfo {author} {\bibfnamefont {L.}~\bibnamefont {Cali{\`o}}},
  \bibinfo {author} {\bibfnamefont {H.}~\bibnamefont {Espin{\'o}s}}, \bibinfo
  {author} {\bibfnamefont {G.}~\bibnamefont {Lavarda}}, \bibinfo {author}
  {\bibfnamefont {T.}~\bibnamefont {Torres}}, \bibinfo {author} {\bibfnamefont
  {J.}~\bibnamefont {Feist}}, \bibinfo {author} {\bibfnamefont {F.~J.}\
  \bibnamefont {Garc{\'\i}a-Vidal}}, \bibinfo {author} {\bibfnamefont
  {G.}~\bibnamefont {Bottari}}, \ and\ \bibinfo {author} {\bibfnamefont
  {H.}~\bibnamefont {M{\'\i}guez}},\ }\href@noop {} {\bibfield  {journal}
  {\bibinfo  {journal} {Solar RRL}\ }\textbf {\bibinfo {volume} {5}},\ \bibinfo
  {pages} {2100308} (\bibinfo {year} {2021})}\BibitemShut {NoStop}%
\bibitem [{\citenamefont {Thomas}\ \emph {et~al.}(2021)\citenamefont {Thomas},
  \citenamefont {Devaux}, \citenamefont {Nagarajan}, \citenamefont {Rogez},
  \citenamefont {Seidel}, \citenamefont {Richard}, \citenamefont {Genet},
  \citenamefont {Drillon},\ and\ \citenamefont {Ebbesen}}]{Thomas2021}%
  \BibitemOpen
  \bibfield  {author} {\bibinfo {author} {\bibfnamefont {A.}~\bibnamefont
  {Thomas}}, \bibinfo {author} {\bibfnamefont {E.}~\bibnamefont {Devaux}},
  \bibinfo {author} {\bibfnamefont {K.}~\bibnamefont {Nagarajan}}, \bibinfo
  {author} {\bibfnamefont {G.}~\bibnamefont {Rogez}}, \bibinfo {author}
  {\bibfnamefont {M.}~\bibnamefont {Seidel}}, \bibinfo {author} {\bibfnamefont
  {F.}~\bibnamefont {Richard}}, \bibinfo {author} {\bibfnamefont
  {C.}~\bibnamefont {Genet}}, \bibinfo {author} {\bibfnamefont
  {M.}~\bibnamefont {Drillon}}, \ and\ \bibinfo {author} {\bibfnamefont
  {T.~W.}\ \bibnamefont {Ebbesen}},\ }\href {\doibase
  10.1021/acs.nanolett.1c00973} {\bibfield  {journal} {\bibinfo  {journal}
  {Nano Letters}\ }\textbf {\bibinfo {volume} {21}},\ \bibinfo {pages} {4365}
  (\bibinfo {year} {2021})},\ \bibinfo {note} {pMID: 33945283}\BibitemShut
  {NoStop}%
\bibitem [{\citenamefont {Chikkaraddy}\ \emph {et~al.}(2016)\citenamefont
  {Chikkaraddy}, \citenamefont {de~Nijs}, \citenamefont {Benz}, \citenamefont
  {Barrow}, \citenamefont {Scherman}, \citenamefont {Rosta}, \citenamefont
  {Demetriadou}, \citenamefont {Fox}, \citenamefont {Hess},\ and\ \citenamefont
  {Baumberg}}]{Chikkaraddy2016}%
  \BibitemOpen
  \bibfield  {author} {\bibinfo {author} {\bibfnamefont {R.}~\bibnamefont
  {Chikkaraddy}}, \bibinfo {author} {\bibfnamefont {B.}~\bibnamefont
  {de~Nijs}}, \bibinfo {author} {\bibfnamefont {F.}~\bibnamefont {Benz}},
  \bibinfo {author} {\bibfnamefont {S.~J.}\ \bibnamefont {Barrow}}, \bibinfo
  {author} {\bibfnamefont {O.~A.}\ \bibnamefont {Scherman}}, \bibinfo {author}
  {\bibfnamefont {E.}~\bibnamefont {Rosta}}, \bibinfo {author} {\bibfnamefont
  {A.}~\bibnamefont {Demetriadou}}, \bibinfo {author} {\bibfnamefont
  {P.}~\bibnamefont {Fox}}, \bibinfo {author} {\bibfnamefont {O.}~\bibnamefont
  {Hess}}, \ and\ \bibinfo {author} {\bibfnamefont {J.~J.}\ \bibnamefont
  {Baumberg}},\ }\href {\doibase 10.1038/nature17974} {\bibfield  {journal}
  {\bibinfo  {journal} {Nature}\ }\textbf {\bibinfo {volume} {535}},\ \bibinfo
  {pages} {127} (\bibinfo {year} {2016})}\BibitemShut {NoStop}%
\bibitem [{\citenamefont {Santhosh}\ \emph {et~al.}(2016)\citenamefont
  {Santhosh}, \citenamefont {Bitton}, \citenamefont {Chuntonov},\ and\
  \citenamefont {Haran}}]{Santhosh2016}%
  \BibitemOpen
  \bibfield  {author} {\bibinfo {author} {\bibfnamefont {K.}~\bibnamefont
  {Santhosh}}, \bibinfo {author} {\bibfnamefont {O.}~\bibnamefont {Bitton}},
  \bibinfo {author} {\bibfnamefont {L.}~\bibnamefont {Chuntonov}}, \ and\
  \bibinfo {author} {\bibfnamefont {G.}~\bibnamefont {Haran}},\ }\href
  {\doibase 10.1038/ncomms11823} {\bibfield  {journal} {\bibinfo  {journal}
  {Nature Communications}\ }\textbf {\bibinfo {volume} {7}},\ \bibinfo {pages}
  {ncomms11823} (\bibinfo {year} {2016})}\BibitemShut {NoStop}%
\bibitem [{\citenamefont {Ojambati}\ \emph {et~al.}(2019)\citenamefont
  {Ojambati}, \citenamefont {Chikkaraddy}, \citenamefont {Deacon},
  \citenamefont {Horton}, \citenamefont {Kos}, \citenamefont {Turek},
  \citenamefont {Keyser},\ and\ \citenamefont {Baumberg}}]{Ojambati2019}%
  \BibitemOpen
  \bibfield  {author} {\bibinfo {author} {\bibfnamefont {O.~S.}\ \bibnamefont
  {Ojambati}}, \bibinfo {author} {\bibfnamefont {R.}~\bibnamefont
  {Chikkaraddy}}, \bibinfo {author} {\bibfnamefont {W.~D.}\ \bibnamefont
  {Deacon}}, \bibinfo {author} {\bibfnamefont {M.}~\bibnamefont {Horton}},
  \bibinfo {author} {\bibfnamefont {D.}~\bibnamefont {Kos}}, \bibinfo {author}
  {\bibfnamefont {V.~A.}\ \bibnamefont {Turek}}, \bibinfo {author}
  {\bibfnamefont {U.~F.}\ \bibnamefont {Keyser}}, \ and\ \bibinfo {author}
  {\bibfnamefont {J.~J.}\ \bibnamefont {Baumberg}},\ }\href {\doibase
  10.1038/s41467-019-08611-5} {\bibfield  {journal} {\bibinfo  {journal}
  {Nature Communications}\ }\textbf {\bibinfo {volume} {10}},\ \bibinfo {pages}
  {1049} (\bibinfo {year} {2019})}\BibitemShut {NoStop}%
\bibitem [{\citenamefont {Park}\ \emph {et~al.}(2019)\citenamefont {Park},
  \citenamefont {May}, \citenamefont {Leng}, \citenamefont {Wang},
  \citenamefont {Kropp}, \citenamefont {Gougousi}, \citenamefont {Pelton},\
  and\ \citenamefont {Raschke}}]{Park2019}%
  \BibitemOpen
  \bibfield  {author} {\bibinfo {author} {\bibfnamefont {K.-D.}\ \bibnamefont
  {Park}}, \bibinfo {author} {\bibfnamefont {M.~A.}\ \bibnamefont {May}},
  \bibinfo {author} {\bibfnamefont {H.}~\bibnamefont {Leng}}, \bibinfo {author}
  {\bibfnamefont {J.}~\bibnamefont {Wang}}, \bibinfo {author} {\bibfnamefont
  {J.~A.}\ \bibnamefont {Kropp}}, \bibinfo {author} {\bibfnamefont
  {T.}~\bibnamefont {Gougousi}}, \bibinfo {author} {\bibfnamefont
  {M.}~\bibnamefont {Pelton}}, \ and\ \bibinfo {author} {\bibfnamefont {M.~B.}\
  \bibnamefont {Raschke}},\ }\href {\doibase 10.1126/sciadv.aav5931} {\bibfield
   {journal} {\bibinfo  {journal} {Science Advances}\ }\textbf {\bibinfo
  {volume} {5}},\ \bibinfo {pages} {eaav5931} (\bibinfo {year}
  {2019})}\BibitemShut {NoStop}%
\bibitem [{\citenamefont {Silva}\ \emph {et~al.}(2020)\citenamefont {Silva},
  \citenamefont {Pino}, \citenamefont {Garc{\'i}a-Vidal},\ and\ \citenamefont
  {Feist}}]{SPGF20}%
  \BibitemOpen
  \bibfield  {author} {\bibinfo {author} {\bibfnamefont {R.~E.~F.}\
  \bibnamefont {Silva}}, \bibinfo {author} {\bibfnamefont {J.~d.}\ \bibnamefont
  {Pino}}, \bibinfo {author} {\bibfnamefont {F.~J.}\ \bibnamefont
  {Garc{\'i}a-Vidal}}, \ and\ \bibinfo {author} {\bibfnamefont
  {J.}~\bibnamefont {Feist}},\ }\href@noop {} {\bibfield  {journal} {\bibinfo
  {journal} {Nature Communications}\ }\textbf {\bibinfo {volume} {11}},\
  \bibinfo {pages} {1423} (\bibinfo {year} {2020})}\BibitemShut {NoStop}%
\bibitem [{\citenamefont {Torres-S{\'a}nchez}\ and\ \citenamefont
  {Feist}(2021)}]{TF21}%
  \BibitemOpen
  \bibfield  {author} {\bibinfo {author} {\bibfnamefont {J.}~\bibnamefont
  {Torres-S{\'a}nchez}}\ and\ \bibinfo {author} {\bibfnamefont
  {J.}~\bibnamefont {Feist}},\ }\href@noop {} {\bibfield  {journal} {\bibinfo
  {journal} {The Journal of Chemical Physics}\ }\textbf {\bibinfo {volume}
  {154}},\ \bibinfo {pages} {014303} (\bibinfo {year} {2021})}\BibitemShut
  {NoStop}%
\bibitem [{\citenamefont {Khurgin}(2015)}]{Khurgin2015}%
  \BibitemOpen
  \bibfield  {author} {\bibinfo {author} {\bibfnamefont {J.~B.}\ \bibnamefont
  {Khurgin}},\ }\href {\doibase 10.1038/nnano.2014.310} {\bibfield  {journal}
  {\bibinfo  {journal} {Nature Nanotechnology}\ }\textbf {\bibinfo {volume}
  {10}},\ \bibinfo {pages} {2} (\bibinfo {year} {2015})}\BibitemShut {NoStop}%
\bibitem [{\citenamefont {Ulusoy}\ and\ \citenamefont {Vendrell}(2020)}]{UV20}%
  \BibitemOpen
  \bibfield  {author} {\bibinfo {author} {\bibfnamefont {I.~S.}\ \bibnamefont
  {Ulusoy}}\ and\ \bibinfo {author} {\bibfnamefont {O.}~\bibnamefont
  {Vendrell}},\ }\href {\doibase 10.1063/5.0011556} {\bibfield  {journal}
  {\bibinfo  {journal} {The Journal of Chemical Physics}\ }\textbf {\bibinfo
  {volume} {153}},\ \bibinfo {pages} {044108} (\bibinfo {year}
  {2020})}\BibitemShut {NoStop}%
\bibitem [{\citenamefont {Du}\ \emph {et~al.}(2021)\citenamefont {Du},
  \citenamefont {Campos-Gonzalez-Angulo},\ and\ \citenamefont
  {Yuen-Zhou}}]{DCY21}%
  \BibitemOpen
  \bibfield  {author} {\bibinfo {author} {\bibfnamefont {M.}~\bibnamefont
  {Du}}, \bibinfo {author} {\bibfnamefont {J.~A.}\ \bibnamefont
  {Campos-Gonzalez-Angulo}}, \ and\ \bibinfo {author} {\bibfnamefont
  {J.}~\bibnamefont {Yuen-Zhou}},\ }\href {\doibase 10.1063/5.0037905}
  {\bibfield  {journal} {\bibinfo  {journal} {The Journal of Chemical Physics}\
  }\textbf {\bibinfo {volume} {154}},\ \bibinfo {pages} {084108} (\bibinfo
  {year} {2021})}\BibitemShut {NoStop}%
\bibitem [{\citenamefont {Antoniou}\ \emph {et~al.}(2020)\citenamefont
  {Antoniou}, \citenamefont {Suchanek}, \citenamefont {Varner},\ and\
  \citenamefont {Foley}}]{ASFVF20}%
  \BibitemOpen
  \bibfield  {author} {\bibinfo {author} {\bibfnamefont {P.}~\bibnamefont
  {Antoniou}}, \bibinfo {author} {\bibfnamefont {F.}~\bibnamefont {Suchanek}},
  \bibinfo {author} {\bibfnamefont {J.~F.}\ \bibnamefont {Varner}}, \ and\
  \bibinfo {author} {\bibfnamefont {J.~J.}\ \bibnamefont {Foley}},\ }\href
  {\doibase 10.1021/acs.jpclett.0c02406} {\bibfield  {journal} {\bibinfo
  {journal} {The Journal of Physical Chemistry Letters}\ }\textbf {\bibinfo
  {volume} {11}},\ \bibinfo {pages} {9063} (\bibinfo {year} {2020})},\ \bibinfo
  {note} {pMID: 33045837}\BibitemShut {NoStop}%
\bibitem [{\citenamefont {Felicetti}\ \emph {et~al.}(2020)\citenamefont
  {Felicetti}, \citenamefont {Fregoni}, \citenamefont {Schnappinger},
  \citenamefont {Reiter}, \citenamefont {de~Vivie-Riedle},\ and\ \citenamefont
  {Feist}}]{FFSRVF20}%
  \BibitemOpen
  \bibfield  {author} {\bibinfo {author} {\bibfnamefont {S.}~\bibnamefont
  {Felicetti}}, \bibinfo {author} {\bibfnamefont {J.}~\bibnamefont {Fregoni}},
  \bibinfo {author} {\bibfnamefont {T.}~\bibnamefont {Schnappinger}}, \bibinfo
  {author} {\bibfnamefont {S.}~\bibnamefont {Reiter}}, \bibinfo {author}
  {\bibfnamefont {R.}~\bibnamefont {de~Vivie-Riedle}}, \ and\ \bibinfo {author}
  {\bibfnamefont {J.}~\bibnamefont {Feist}},\ }\href {\doibase
  10.1021/acs.jpclett.0c02236} {\bibfield  {journal} {\bibinfo  {journal} {The
  Journal of Physical Chemistry Letters}\ }\textbf {\bibinfo {volume} {11}},\
  \bibinfo {pages} {8810} (\bibinfo {year} {2020})},\ \bibinfo {note} {pMID:
  32914984}\BibitemShut {NoStop}%
\bibitem [{\citenamefont {Tokatly}(2013)}]{T13}%
  \BibitemOpen
  \bibfield  {author} {\bibinfo {author} {\bibfnamefont {I.~V.}\ \bibnamefont
  {Tokatly}},\ }\href@noop {} {\bibfield  {journal} {\bibinfo  {journal} {Phys.
  Rev. Lett.}\ }\textbf {\bibinfo {volume} {110}},\ \bibinfo {pages} {233001}
  (\bibinfo {year} {2013})}\BibitemShut {NoStop}%
\bibitem [{\citenamefont {Ruggenthaler}\ \emph {et~al.}(2014)\citenamefont
  {Ruggenthaler}, \citenamefont {Flick}, \citenamefont {Pellegrini},
  \citenamefont {Appel}, \citenamefont {Tokatly},\ and\ \citenamefont
  {Rubio}}]{RFPATR14}%
  \BibitemOpen
  \bibfield  {author} {\bibinfo {author} {\bibfnamefont {M.}~\bibnamefont
  {Ruggenthaler}}, \bibinfo {author} {\bibfnamefont {J.}~\bibnamefont {Flick}},
  \bibinfo {author} {\bibfnamefont {C.}~\bibnamefont {Pellegrini}}, \bibinfo
  {author} {\bibfnamefont {H.}~\bibnamefont {Appel}}, \bibinfo {author}
  {\bibfnamefont {I.~V.}\ \bibnamefont {Tokatly}}, \ and\ \bibinfo {author}
  {\bibfnamefont {A.}~\bibnamefont {Rubio}},\ }\href@noop {} {\bibfield
  {journal} {\bibinfo  {journal} {Phys. Rev. A}\ }\textbf {\bibinfo {volume}
  {90}},\ \bibinfo {pages} {012508} (\bibinfo {year} {2014})}\BibitemShut
  {NoStop}%
\bibitem [{\citenamefont {Ruggenthaler}\ \emph {et~al.}(2018)\citenamefont
  {Ruggenthaler}, \citenamefont {Tancogne-Dejean}, \citenamefont {Flick},
  \citenamefont {Appel},\ and\ \citenamefont {Rubio}}]{RTFAR18}%
  \BibitemOpen
  \bibfield  {author} {\bibinfo {author} {\bibfnamefont {M.}~\bibnamefont
  {Ruggenthaler}}, \bibinfo {author} {\bibfnamefont {N.}~\bibnamefont
  {Tancogne-Dejean}}, \bibinfo {author} {\bibfnamefont {J.}~\bibnamefont
  {Flick}}, \bibinfo {author} {\bibfnamefont {H.}~\bibnamefont {Appel}}, \ and\
  \bibinfo {author} {\bibfnamefont {A.}~\bibnamefont {Rubio}},\ }\href@noop {}
  {\bibfield  {journal} {\bibinfo  {journal} {Nature Reviews Chemistry}\
  }\textbf {\bibinfo {volume} {2}},\ \bibinfo {pages} {0118} (\bibinfo {year}
  {2018})}\BibitemShut {NoStop}%
\bibitem [{\citenamefont {S{\'a}nchez-Barquilla}\ \emph
  {et~al.}(2020)\citenamefont {S{\'a}nchez-Barquilla}, \citenamefont {Silva},\
  and\ \citenamefont {Feist}}]{SSF20}%
  \BibitemOpen
  \bibfield  {author} {\bibinfo {author} {\bibfnamefont {M.}~\bibnamefont
  {S{\'a}nchez-Barquilla}}, \bibinfo {author} {\bibfnamefont {R.}~\bibnamefont
  {Silva}}, \ and\ \bibinfo {author} {\bibfnamefont {J.}~\bibnamefont
  {Feist}},\ }\href@noop {} {\bibfield  {journal} {\bibinfo  {journal} {The
  Journal of chemical physics}\ }\textbf {\bibinfo {volume} {152}},\ \bibinfo
  {pages} {034108} (\bibinfo {year} {2020})}\BibitemShut {NoStop}%
\bibitem [{\citenamefont {del Pino}\ \emph {et~al.}(2018)\citenamefont {del
  Pino}, \citenamefont {Schr{\"o}der}, \citenamefont {Chin}, \citenamefont
  {Feist},\ and\ \citenamefont {Garcia-Vidal}}]{PSCFG18}%
  \BibitemOpen
  \bibfield  {author} {\bibinfo {author} {\bibfnamefont {J.}~\bibnamefont {del
  Pino}}, \bibinfo {author} {\bibfnamefont {F.~A. Y.~N.}\ \bibnamefont
  {Schr{\"o}der}}, \bibinfo {author} {\bibfnamefont {A.~W.}\ \bibnamefont
  {Chin}}, \bibinfo {author} {\bibfnamefont {J.}~\bibnamefont {Feist}}, \ and\
  \bibinfo {author} {\bibfnamefont {F.~J.}\ \bibnamefont {Garcia-Vidal}},\
  }\href {\doibase 10.1103/PhysRevLett.121.227401} {\bibfield  {journal}
  {\bibinfo  {journal} {Phys. Rev. Lett.}\ }\textbf {\bibinfo {volume} {121}},\
  \bibinfo {pages} {227401} (\bibinfo {year} {2018})}\BibitemShut {NoStop}%
\bibitem [{\citenamefont {Franke}\ \emph {et~al.}(2019)\citenamefont {Franke},
  \citenamefont {Hughes}, \citenamefont {Dezfouli}, \citenamefont {Kristensen},
  \citenamefont {Busch}, \citenamefont {Knorr},\ and\ \citenamefont
  {Richter}}]{FHDKBKR19}%
  \BibitemOpen
  \bibfield  {author} {\bibinfo {author} {\bibfnamefont {S.}~\bibnamefont
  {Franke}}, \bibinfo {author} {\bibfnamefont {S.}~\bibnamefont {Hughes}},
  \bibinfo {author} {\bibfnamefont {M.~K.}\ \bibnamefont {Dezfouli}}, \bibinfo
  {author} {\bibfnamefont {P.~T.}\ \bibnamefont {Kristensen}}, \bibinfo
  {author} {\bibfnamefont {K.}~\bibnamefont {Busch}}, \bibinfo {author}
  {\bibfnamefont {A.}~\bibnamefont {Knorr}}, \ and\ \bibinfo {author}
  {\bibfnamefont {M.}~\bibnamefont {Richter}},\ }\href {\doibase
  10.1103/PhysRevLett.122.213901} {\bibfield  {journal} {\bibinfo  {journal}
  {Phys. Rev. Lett.}\ }\textbf {\bibinfo {volume} {122}},\ \bibinfo {pages}
  {213901} (\bibinfo {year} {2019})}\BibitemShut {NoStop}%
\bibitem [{\citenamefont {Hoffmann}\ \emph
  {et~al.}(2019{\natexlab{a}})\citenamefont {Hoffmann}, \citenamefont
  {Sch{\"a}fer}, \citenamefont {Rubio}, \citenamefont {Kelly},\ and\
  \citenamefont {Appel}}]{HSRKA19}%
  \BibitemOpen
  \bibfield  {author} {\bibinfo {author} {\bibfnamefont {N.~M.}\ \bibnamefont
  {Hoffmann}}, \bibinfo {author} {\bibfnamefont {C.}~\bibnamefont
  {Sch{\"a}fer}}, \bibinfo {author} {\bibfnamefont {A.}~\bibnamefont {Rubio}},
  \bibinfo {author} {\bibfnamefont {A.}~\bibnamefont {Kelly}}, \ and\ \bibinfo
  {author} {\bibfnamefont {H.}~\bibnamefont {Appel}},\ }\href@noop {}
  {\bibfield  {journal} {\bibinfo  {journal} {Phys. Rev. A}\ }\textbf {\bibinfo
  {volume} {99}},\ \bibinfo {pages} {063819} (\bibinfo {year}
  {2019}{\natexlab{a}})}\BibitemShut {NoStop}%
\bibitem [{\citenamefont {Hoffmann}\ \emph
  {et~al.}(2019{\natexlab{b}})\citenamefont {Hoffmann}, \citenamefont
  {Sch{\"a}fer}, \citenamefont {S{\"a}kkinen}, \citenamefont {Rubio},
  \citenamefont {Appel},\ and\ \citenamefont {Kelly}}]{HSSRAK19}%
  \BibitemOpen
  \bibfield  {author} {\bibinfo {author} {\bibfnamefont {N.~M.}\ \bibnamefont
  {Hoffmann}}, \bibinfo {author} {\bibfnamefont {C.}~\bibnamefont
  {Sch{\"a}fer}}, \bibinfo {author} {\bibfnamefont {N.}~\bibnamefont
  {S{\"a}kkinen}}, \bibinfo {author} {\bibfnamefont {A.}~\bibnamefont {Rubio}},
  \bibinfo {author} {\bibfnamefont {H.}~\bibnamefont {Appel}}, \ and\ \bibinfo
  {author} {\bibfnamefont {A.}~\bibnamefont {Kelly}},\ }\href@noop {}
  {\bibfield  {journal} {\bibinfo  {journal} {The Journal of Chemical Physics}\
  }\textbf {\bibinfo {volume} {151}},\ \bibinfo {pages} {244113} (\bibinfo
  {year} {2019}{\natexlab{b}})}\BibitemShut {NoStop}%
\bibitem [{\citenamefont {Hoffmann}\ \emph {et~al.}(2020)\citenamefont
  {Hoffmann}, \citenamefont {Lacombe}, \citenamefont {Rubio},\ and\
  \citenamefont {Maitra}}]{HLRM20}%
  \BibitemOpen
  \bibfield  {author} {\bibinfo {author} {\bibfnamefont {N.~M.}\ \bibnamefont
  {Hoffmann}}, \bibinfo {author} {\bibfnamefont {L.}~\bibnamefont {Lacombe}},
  \bibinfo {author} {\bibfnamefont {A.}~\bibnamefont {Rubio}}, \ and\ \bibinfo
  {author} {\bibfnamefont {N.~T.}\ \bibnamefont {Maitra}},\ }\href@noop {}
  {\bibfield  {journal} {\bibinfo  {journal} {The Journal of Chemical Physics}\
  }\textbf {\bibinfo {volume} {153}},\ \bibinfo {pages} {104103} (\bibinfo
  {year} {2020})}\BibitemShut {NoStop}%
\bibitem [{\citenamefont {Li}\ \emph {et~al.}(2020)\citenamefont {Li},
  \citenamefont {Subotnik},\ and\ \citenamefont {Nitzan}}]{LSN20}%
  \BibitemOpen
  \bibfield  {author} {\bibinfo {author} {\bibfnamefont {T.~E.}\ \bibnamefont
  {Li}}, \bibinfo {author} {\bibfnamefont {J.~E.}\ \bibnamefont {Subotnik}}, \
  and\ \bibinfo {author} {\bibfnamefont {A.}~\bibnamefont {Nitzan}},\
  }\href@noop {} {\bibfield  {journal} {\bibinfo  {journal} {Proceedings of the
  National Academy of Sciences}\ }\textbf {\bibinfo {volume} {117}},\ \bibinfo
  {pages} {18324} (\bibinfo {year} {2020})}\BibitemShut {NoStop}%
\bibitem [{\citenamefont {Li}\ \emph {et~al.}(2021)\citenamefont {Li},
  \citenamefont {Mandal},\ and\ \citenamefont {Huo}}]{LMH21}%
  \BibitemOpen
  \bibfield  {author} {\bibinfo {author} {\bibfnamefont {X.}~\bibnamefont
  {Li}}, \bibinfo {author} {\bibfnamefont {A.}~\bibnamefont {Mandal}}, \ and\
  \bibinfo {author} {\bibfnamefont {P.}~\bibnamefont {Huo}},\ }\href@noop {}
  {\bibfield  {journal} {\bibinfo  {journal} {Nature Communications}\ }\textbf
  {\bibinfo {volume} {12}},\ \bibinfo {pages} {1315} (\bibinfo {year}
  {2021})}\BibitemShut {NoStop}%
\bibitem [{\citenamefont {Heller}(1976)}]{H76}%
  \BibitemOpen
  \bibfield  {author} {\bibinfo {author} {\bibfnamefont {E.~J.}\ \bibnamefont
  {Heller}},\ }\href@noop {} {\bibfield  {journal} {\bibinfo  {journal} {The
  Journal of Chemical Physics}\ }\textbf {\bibinfo {volume} {65}},\ \bibinfo
  {pages} {1289} (\bibinfo {year} {1976})}\BibitemShut {NoStop}%
\bibitem [{\citenamefont {Abedi}\ \emph {et~al.}(2010)\citenamefont {Abedi},
  \citenamefont {Maitra},\ and\ \citenamefont {Gross}}]{AMG10}%
  \BibitemOpen
  \bibfield  {author} {\bibinfo {author} {\bibfnamefont {A.}~\bibnamefont
  {Abedi}}, \bibinfo {author} {\bibfnamefont {N.~T.}\ \bibnamefont {Maitra}}, \
  and\ \bibinfo {author} {\bibfnamefont {E.~K.~U.}\ \bibnamefont {Gross}},\
  }\href@noop {} {\bibfield  {journal} {\bibinfo  {journal} {Phys. Rev. Lett.}\
  }\textbf {\bibinfo {volume} {105}},\ \bibinfo {pages} {123002} (\bibinfo
  {year} {2010})}\BibitemShut {NoStop}%
\bibitem [{\citenamefont {Abedi}\ \emph {et~al.}(2012)\citenamefont {Abedi},
  \citenamefont {Maitra},\ and\ \citenamefont {Gross}}]{AMG12}%
  \BibitemOpen
  \bibfield  {author} {\bibinfo {author} {\bibfnamefont {A.}~\bibnamefont
  {Abedi}}, \bibinfo {author} {\bibfnamefont {N.~T.}\ \bibnamefont {Maitra}}, \
  and\ \bibinfo {author} {\bibfnamefont {E.~K.~U.}\ \bibnamefont {Gross}},\
  }\href@noop {} {\bibfield  {journal} {\bibinfo  {journal} {The Journal of
  Chemical Physics}\ }\textbf {\bibinfo {volume} {137}},\ \bibinfo {pages}
  {22A530} (\bibinfo {year} {2012})}\BibitemShut {NoStop}%
\bibitem [{\citenamefont {Hoffmann}\ \emph {et~al.}(2018)\citenamefont
  {Hoffmann}, \citenamefont {Appel}, \citenamefont {Rubio},\ and\ \citenamefont
  {Maitra}}]{HARM18}%
  \BibitemOpen
  \bibfield  {author} {\bibinfo {author} {\bibfnamefont {N.~M.}\ \bibnamefont
  {Hoffmann}}, \bibinfo {author} {\bibfnamefont {H.}~\bibnamefont {Appel}},
  \bibinfo {author} {\bibfnamefont {A.}~\bibnamefont {Rubio}}, \ and\ \bibinfo
  {author} {\bibfnamefont {N.~T.}\ \bibnamefont {Maitra}},\ }\href@noop {}
  {\bibfield  {journal} {\bibinfo  {journal} {The European Physical Journal B}\
  }\textbf {\bibinfo {volume} {91}},\ \bibinfo {pages} {180} (\bibinfo {year}
  {2018})}\BibitemShut {NoStop}%
\bibitem [{\citenamefont {Agostini}\ and\ \citenamefont {Gross}(2021)}]{AG21}%
  \BibitemOpen
  \bibfield  {author} {\bibinfo {author} {\bibfnamefont {F.}~\bibnamefont
  {Agostini}}\ and\ \bibinfo {author} {\bibfnamefont {E.~K.~U.}\ \bibnamefont
  {Gross}},\ }\href@noop {} {\bibfield  {journal} {\bibinfo  {journal} {The
  European Physical Journal B}\ }\textbf {\bibinfo {volume} {94}},\ \bibinfo
  {pages} {179} (\bibinfo {year} {2021})}\BibitemShut {NoStop}%
\bibitem [{\citenamefont {Lacombe}\ \emph {et~al.}(2019)\citenamefont
  {Lacombe}, \citenamefont {Hoffmann},\ and\ \citenamefont {Maitra}}]{LHM19}%
  \BibitemOpen
  \bibfield  {author} {\bibinfo {author} {\bibfnamefont {L.}~\bibnamefont
  {Lacombe}}, \bibinfo {author} {\bibfnamefont {N.~M.}\ \bibnamefont
  {Hoffmann}}, \ and\ \bibinfo {author} {\bibfnamefont {N.~T.}\ \bibnamefont
  {Maitra}},\ }\href@noop {} {\bibfield  {journal} {\bibinfo  {journal} {Phys.
  Rev. Lett.}\ }\textbf {\bibinfo {volume} {123}},\ \bibinfo {pages} {083201}
  (\bibinfo {year} {2019})}\BibitemShut {NoStop}%
\bibitem [{\citenamefont {Martinez}\ \emph {et~al.}(2021)\citenamefont
  {Martinez}, \citenamefont {Rosenzweig}, \citenamefont {Hoffmann},
  \citenamefont {Lacombe},\ and\ \citenamefont {Maitra}}]{MRHLM21}%
  \BibitemOpen
  \bibfield  {author} {\bibinfo {author} {\bibfnamefont {P.}~\bibnamefont
  {Martinez}}, \bibinfo {author} {\bibfnamefont {B.}~\bibnamefont
  {Rosenzweig}}, \bibinfo {author} {\bibfnamefont {N.~M.}\ \bibnamefont
  {Hoffmann}}, \bibinfo {author} {\bibfnamefont {L.}~\bibnamefont {Lacombe}}, \
  and\ \bibinfo {author} {\bibfnamefont {N.~T.}\ \bibnamefont {Maitra}},\
  }\href@noop {} {\bibfield  {journal} {\bibinfo  {journal} {The Journal of
  Chemical Physics}\ }\textbf {\bibinfo {volume} {154}},\ \bibinfo {pages}
  {014102} (\bibinfo {year} {2021})}\BibitemShut {NoStop}%
\bibitem [{\citenamefont {Abedi}\ \emph {et~al.}(2018)\citenamefont {Abedi},
  \citenamefont {Khosravi},\ and\ \citenamefont {Tokatly}}]{AKT18}%
  \BibitemOpen
  \bibfield  {author} {\bibinfo {author} {\bibfnamefont {A.}~\bibnamefont
  {Abedi}}, \bibinfo {author} {\bibfnamefont {E.}~\bibnamefont {Khosravi}}, \
  and\ \bibinfo {author} {\bibfnamefont {I.~V.}\ \bibnamefont {Tokatly}},\
  }\href@noop {} {\bibfield  {journal} {\bibinfo  {journal} {The European
  Physical Journal B}\ }\textbf {\bibinfo {volume} {91}},\ \bibinfo {pages}
  {194} (\bibinfo {year} {2018})}\BibitemShut {NoStop}%
\bibitem [{\citenamefont {Cohen-Tannoudji}\ \emph {et~al.}()\citenamefont
  {Cohen-Tannoudji}, \citenamefont {Dupont-Roc},\ and\ \citenamefont
  {Grynberg}}]{CohenTannoudjiBook}%
  \BibitemOpen
  \bibfield  {author} {\bibinfo {author} {\bibfnamefont {C.}~\bibnamefont
  {Cohen-Tannoudji}}, \bibinfo {author} {\bibfnamefont {J.}~\bibnamefont
  {Dupont-Roc}}, \ and\ \bibinfo {author} {\bibfnamefont {G.}~\bibnamefont
  {Grynberg}},\ }\enquote {\bibinfo {title} {Quantum electrodynamics in the
  coulomb gauge},}\ in\ \href@noop {} {\emph {\bibinfo {booktitle} {Photons and
  Atoms}}}\ (\bibinfo  {publisher} {John Wiley \& Sons, Ltd})\ Chap.~\bibinfo
  {chapter} {3}, pp.\ \bibinfo {pages} {169--252}\BibitemShut {NoStop}%
\bibitem [{\citenamefont {Tully}(1998)}]{T98}%
  \BibitemOpen
  \bibfield  {author} {\bibinfo {author} {\bibfnamefont {J.~C.}\ \bibnamefont
  {Tully}},\ }\href@noop {} {\bibfield  {journal} {\bibinfo  {journal} {Faraday
  Discuss.}\ }\textbf {\bibinfo {volume} {110}},\ \bibinfo {pages} {407}
  (\bibinfo {year} {1998})}\BibitemShut {NoStop}%
\bibitem [{\citenamefont {Meyer}\ and\ \citenamefont {Miller}(1980)}]{MM80}%
  \BibitemOpen
  \bibfield  {author} {\bibinfo {author} {\bibfnamefont {H.}~\bibnamefont
  {Meyer}}\ and\ \bibinfo {author} {\bibfnamefont {W.~H.}\ \bibnamefont
  {Miller}},\ }\href {\doibase 10.1063/1.439462} {\bibfield  {journal}
  {\bibinfo  {journal} {The Journal of Chemical Physics}\ }\textbf {\bibinfo
  {volume} {72}},\ \bibinfo {pages} {2272} (\bibinfo {year}
  {1980})}\BibitemShut {NoStop}%
\bibitem [{\citenamefont {Agostini}\ \emph {et~al.}(2015)\citenamefont
  {Agostini}, \citenamefont {Abedi}, \citenamefont {Suzuki}, \citenamefont
  {Min}, \citenamefont {Maitra},\ and\ \citenamefont {Gross}}]{AASMMG15}%
  \BibitemOpen
  \bibfield  {author} {\bibinfo {author} {\bibfnamefont {F.}~\bibnamefont
  {Agostini}}, \bibinfo {author} {\bibfnamefont {A.}~\bibnamefont {Abedi}},
  \bibinfo {author} {\bibfnamefont {Y.}~\bibnamefont {Suzuki}}, \bibinfo
  {author} {\bibfnamefont {S.~K.}\ \bibnamefont {Min}}, \bibinfo {author}
  {\bibfnamefont {N.~T.}\ \bibnamefont {Maitra}}, \ and\ \bibinfo {author}
  {\bibfnamefont {E.~K.~U.}\ \bibnamefont {Gross}},\ }\href@noop {} {\bibfield
  {journal} {\bibinfo  {journal} {J. Chem. Phys.}\ }\textbf {\bibinfo {volume}
  {142}},\ \bibinfo {pages} {084303} (\bibinfo {year} {2015})}\BibitemShut
  {NoStop}%
\bibitem [{\citenamefont {Min}\ \emph {et~al.}(2015)\citenamefont {Min},
  \citenamefont {Agostini},\ and\ \citenamefont {Gross}}]{MAG15}%
  \BibitemOpen
  \bibfield  {author} {\bibinfo {author} {\bibfnamefont {S.~K.}\ \bibnamefont
  {Min}}, \bibinfo {author} {\bibfnamefont {F.}~\bibnamefont {Agostini}}, \
  and\ \bibinfo {author} {\bibfnamefont {E.~K.~U.}\ \bibnamefont {Gross}},\
  }\href@noop {} {\bibfield  {journal} {\bibinfo  {journal} {Phys. Rev. Lett.}\
  }\textbf {\bibinfo {volume} {115}},\ \bibinfo {pages} {073001} (\bibinfo
  {year} {2015})}\BibitemShut {NoStop}%
\bibitem [{\citenamefont {Agostini}\ \emph {et~al.}(2016)\citenamefont
  {Agostini}, \citenamefont {Min}, \citenamefont {Abedi},\ and\ \citenamefont
  {Gross}}]{AMAG16}%
  \BibitemOpen
  \bibfield  {author} {\bibinfo {author} {\bibfnamefont {F.}~\bibnamefont
  {Agostini}}, \bibinfo {author} {\bibfnamefont {S.~K.}\ \bibnamefont {Min}},
  \bibinfo {author} {\bibfnamefont {A.}~\bibnamefont {Abedi}}, \ and\ \bibinfo
  {author} {\bibfnamefont {E.~K.~U.}\ \bibnamefont {Gross}},\ }\href@noop {}
  {\bibfield  {journal} {\bibinfo  {journal} {Journal of Chemical Theory and
  Computation}\ }\textbf {\bibinfo {volume} {12}},\ \bibinfo {pages} {2127}
  (\bibinfo {year} {2016})}\BibitemShut {NoStop}%
\bibitem [{\citenamefont {Min}\ \emph {et~al.}(2017)\citenamefont {Min},
  \citenamefont {Agostini}, \citenamefont {Tavernelli},\ and\ \citenamefont
  {Gross}}]{MATG17}%
  \BibitemOpen
  \bibfield  {author} {\bibinfo {author} {\bibfnamefont {S.~K.}\ \bibnamefont
  {Min}}, \bibinfo {author} {\bibfnamefont {F.}~\bibnamefont {Agostini}},
  \bibinfo {author} {\bibfnamefont {I.}~\bibnamefont {Tavernelli}}, \ and\
  \bibinfo {author} {\bibfnamefont {E.~K.~U.}\ \bibnamefont {Gross}},\
  }\href@noop {} {\bibfield  {journal} {\bibinfo  {journal} {The Journal of
  Physical Chemistry Letters}\ }\textbf {\bibinfo {volume} {8}},\ \bibinfo
  {pages} {3048} (\bibinfo {year} {2017})}\BibitemShut {NoStop}%
\bibitem [{\citenamefont {Ha}\ \emph {et~al.}(2018)\citenamefont {Ha},
  \citenamefont {Lee},\ and\ \citenamefont {Min}}]{HLM18}%
  \BibitemOpen
  \bibfield  {author} {\bibinfo {author} {\bibfnamefont {J.-K.}\ \bibnamefont
  {Ha}}, \bibinfo {author} {\bibfnamefont {I.~S.}\ \bibnamefont {Lee}}, \ and\
  \bibinfo {author} {\bibfnamefont {S.~K.}\ \bibnamefont {Min}},\ }\href@noop
  {} {\bibfield  {journal} {\bibinfo  {journal} {J. Phys. Chem. Lett.}\
  }\textbf {\bibinfo {volume} {9}},\ \bibinfo {pages} {1097} (\bibinfo {year}
  {2018})}\BibitemShut {NoStop}%
\bibitem [{\citenamefont {Vindel-Zandbergen}\ \emph {et~al.}(2021)\citenamefont
  {Vindel-Zandbergen}, \citenamefont {Ibele}, \citenamefont {Ha}, \citenamefont
  {Min}, \citenamefont {Curchod},\ and\ \citenamefont {Maitra}}]{VIHMCM21}%
  \BibitemOpen
  \bibfield  {author} {\bibinfo {author} {\bibfnamefont {P.}~\bibnamefont
  {Vindel-Zandbergen}}, \bibinfo {author} {\bibfnamefont {L.~M.}\ \bibnamefont
  {Ibele}}, \bibinfo {author} {\bibfnamefont {J.-K.}\ \bibnamefont {Ha}},
  \bibinfo {author} {\bibfnamefont {S.~K.}\ \bibnamefont {Min}}, \bibinfo
  {author} {\bibfnamefont {B.~F.~E.}\ \bibnamefont {Curchod}}, \ and\ \bibinfo
  {author} {\bibfnamefont {N.~T.}\ \bibnamefont {Maitra}},\ }\href@noop {}
  {\bibfield  {journal} {\bibinfo  {journal} {Journal of Chemical Theory and
  Computation}\ }\textbf {\bibinfo {volume} {17}},\ \bibinfo {pages} {3852}
  (\bibinfo {year} {2021})}\BibitemShut {NoStop}%
\end{thebibliography}%

\end{document}